\documentclass[12pt,preprint]{aastex}

\begin{document}

\title{An infrared survey of brightest cluster galaxies: Paper I}
\author{Alice C. Quillen, Nicholas Zufelt, Jaehong Park}
\affil{Department of Physics and Astronomy,
  University of Rochester, Rochester, NY 14627}
\email{aquillen@pas.rochester.edu, zufelt72@potsdam.edu,
jaehong@pas.rochester.edu}
\author{Christopher P. O'Dea, Stefi A. Baum, George Privon, Jacob Noel-Storr}
\affil{Department of Physics, Rochester Institute of Technology,
  84 Lomb Memorial Drive, Rochester, NY 14623-5603}
\email{odea@cis.rit.edu, baum@cis.rit.edu, gcp1035@cis.rit.edu, jake@cis.rit.edu}
\author{Alastair Edge}
\affil{Institute for Computational Cosmology, Department of Physics,
  Durham University, Durham DH1 3LE}
\email{alastair.edge@durham.ac.uk}
\author{Helen Russell, Andy Fabian}
\affil{Institute of Astronomy, Madingley Rd., Cambridge,
  CB30HA, UK} 
\email{hrr27@ast.cam.ac.uk}
\author{Megan Donahue}
\affil{Michigan State University, Physics and Astronomy Dept.,
  East Lansing, MI 48824-2320} 
\email{donahue@pa.msu.edu}
\author{Joel N. Bregman}
\affil{University of Michigan, Physics  Dept.,
Ann Arbor, MI 48109} 
\email{jbregman@umich.edu}
\author{Brian R. McNamara}
\affil{Department of Physics \& Astronomy,
University of Waterloo, 200 University Avenue West,
Waterloo, Ontario, Canada N2L 3G1} \email{mcnamara@uwaterloo.ca}
\author {\&}
\author{Craig L. Sarazin}
\affil{Department of Astronomy, University of Virginia, P.O. Box 400325, Charlottesville, VA 22904-4325} 
\email{sarazin@virginia.edu}


\begin{abstract}
We report on an imaging survey with the {\it Spitzer Space Telescope }
of 62 brightest cluster galaxies with optical line emission.
These galaxies are located in
the cores of X-ray luminous clusters selected from the ROSAT 
All-Sky Survey.
We find that about half of these sources have a sign of excess infrared
emission;  22 objects out of 62 are detected at 70 $\mu$m, 18 have
8 to 5.8 $\mu$m flux ratios above 1.0 and 28 have 24 to
8 $\mu$m flux ratios above 1.0.  Altogether 
35 of 62 objects in our survey exhibit at least
one of these signs of infrared excess.
Four galaxies with infrared excesses have
a  4.5/3.6 $\mu$m flux ratio indicating the presence
of hot dust, and/or an unresolved nucleus at 8 $\mu$m.
Three of these have high measured $[$OIII$]$(5007\AA)/H$\beta$
flux ratios suggesting that these four,
Abell 1068, Abell 2146, and Zwicky 2089, and R0821+07,
host dusty active galactic nuclei (AGNs).
9 objects (including the four hosting dusty AGNs) have infrared luminosities
greater than $10^{11} L_\odot$ and so can be classified as luminous infrared
galaxies (LIRGs).
Excluding the four systems hosting dusty AGNs, 
the excess mid-infrared emission in
the remaining brightest cluster galaxies is likely related to star formation.
\end{abstract}

\keywords{stars: formation 
-- galaxies: clusters: 
general -- galaxies: active -- 
galaxies: elliptical and lenticular, 
cD --  cooling flows -- infrared: galaxies }

\section{Introduction}

X-ray observations of many galaxy clusters show a density increase
and a temperature decrease toward the centers of these systems, implying
that gas should be cooling at rates of a few to 1000~$M_\odot$~yr$^{-1}$
(e.g., \citealt{allen00}).
A number of studies have found evidence for cooled gas
and star formation in galaxy clusters, but at a rate below that corresponding
to the cooling rate predicted from the X-ray observations
(e.g., \citealt{mcnamara89,cardiel98,crawford99,bregman06}).
Although mechanisms to counteract radiative cooling of
the intracluster medium have been identified 
(e.g., \citealt{birzan04,rafferty06,dunn06,peterson06,mcnamara07}),
these energy sources do not seem to be sufficiently efficient
to shut off cooling altogether.  Up to $10^{11} M_\odot$ molecular
gas is detected in BCGs with the largest mass deposition rates 
(e.g., \citealt{edge02,salome03}).
Most cooling flow systems exhibit optical emission-lines such
as H$\alpha$ (\citealt{heckman81,Hu85}).  
Emission line spectroscopy suggests that some
of the emission is due to photoionization and heating by hot stars
\citep{voit97}.
The gulf between the spectral limits on the condensation rates
and the sink of cooling material, the primary objection to cooling flows,
has narrowed dramatically.  In some systems, star formation rates
are close to the X-ray and far UV estimates of the rates of gas condensation
\citep{hicks05,rafferty06,mcnamara04}.

To probe the efficiency of cooling and star formation in cluster galaxies,
we aim to provide more direct measurements of the star formation
rate in central  cluster galaxies using broad band
images obtained with the {\it Spitzer Space Telescope} (SST).
Recent work using SST observations of a few 
brightest cluster galaxies (BCGs) includes the study by
\citet{egami06} finding that the majority of BCGs are not
infrared luminous.  This study of 11 galaxies suggested that
star formation was responsible for non-AGN excess infrared emission
in 3 BCGs and that star formation was taking place primarily
in objects with high X-ray luminosity.  Two of the objects studied
by \citet{egami06} have estimated infrared luminosities classifying them as
Luminous Infrared Galaxies (LIRGs),
$10^{11} < L_{\rm IR} < 10^{12} L_\odot$.
Star formation rates for the infrared luminous objects were approximately
10 times lower than the mass deposition rates.
While the galaxies studied by \citet{egami06} are moderately distant
($0.15< z<0.5$), \citet{donahue07} found
that the BCG in the nearby ($z=0.0852$) cluster Abell 2597 can
also be classified as a LIRG.

In this paper we build on the previous studies of a small number of BCGs
\citep{egami06,donahue07} with an imaging survey of 62 BCGs using the
Spitzer Space Telescope (SST). 
In this paper we describe the sample and survey data and briefly discuss
the morphology seen in the images.  Using aperture photometry
we assemble spectral energy distributions (SEDs) of the central regions of
the galaxies and identify the sources that have infrared excesses
and estimate their infrared luminosities.
Using infrared colors and morphology (whether a central source
was red and unresolved) and optical emission line ratios we differentiate
between sources with infrared emission consistent with contribution
from a dusty AGN
and those likely dominated by star formation. Comparison data
are assembled to search for correlations between infrared luminosities
and other cluster properties; however, the actual comparisons are done
in a companion paper (paper II; \citealt{odea07}).
In this paper all luminosities
have been corrected or computed to be consistent with a Hubble constant
$H_0 = 70$~Mpc$^{-1}$~km~s$^{-1}$ and a concordant cosmology
($\Omega_M=0.3$ and flat).

\section{Images}

\subsection{Sample}

The sample was chosen to investigate the impact of optical
emission lines, known to be a good indicator of a cool core cluster
\citep{edge92,peres98}, on the mid-infrared emission
of BCGs in a representative sample of clusters. The sample was
principally selected from the comprehensive optical spectral survey of
BCGs in 215 ROSAT-selected clusters of \citet{crawford99}.
The Crawford et al. study includes 177 clusters (or 87\%)
from the Brightest Cluster Sample (BCS, \citealt{ebeling98}),
17 (17\%) from the extended Brightest Cluster Sample 
(BCS, \citealt{ebeling00}), 
11 (2\%) of the ROSAT-ESO Flux Limited X-ray (REFLEX) cluster
sample \citep{bohringer04} and 10 clusters below
the flux limit from these 3 samples ($<3 \times 10^{-12}$
erg~cm$^{-2}$~s$^{-1}$ 0.1-2.4~keV) or at low galactic latitude
($|b|<20^\circ$). Optical emission lines are detected in 27\% of
the Crawford et al. sample and we select all BCGs detected
brighter than 0.5$\times 10^{-15}$ erg~cm$^{-2}$~s$^{-1}$
extracted from a 1.3$''$ slit
in either H$\alpha$ or [NII] from the BCS, eBCS and REFLEX
samples.  This flux selection is a factor of a few above
the detection limit of this study so allows for
any variation in depth due to observing conditions
and continuum strength.
Excluding three objects where the lines were affected
by either atmospheric absorption (A671), high declination
(A2294) or a badly placed cosmic ray (RXJ1750.2$+$3505),
this gives a total of sample of 64 BCGs. To this
sample we add the 6 BCS BCGs that are known to exhibit
line emission but were not observed by Crawford et al. to
ensure that we have complete sampling of the BCS line
emitters. Of this sample of 70, 16 have existing GTO or
GO Spitzer observations, e.g. M87/3C274, NGC 383/3C31 and NGC7237/3C442A
in the study of powerful FRI radio galaxies by Birkinshaw
and A1835, Zw3146 and Zw7160
in the sample by \citet{egami06}, leaving 54 to be observed
with imaging by the SST.

For completeness and to include clusters in both hemispheres,
we have supplemented the remaining
total of 54 line emitting BCGs from the study by Crawford et al.
with four well-studied BCGs from the \citet{edge90}
all-sky sample (A85, A3112, A4059 and PKS0745-191),
Abell S1101 (frequently referred to as Sersic159-01) from REFLEX
and Z348 from the eBCS which shows very strong emission lines in
its spectrum in the SDSS. One BCG from a cluster just below our
nominal X-ray flux limit was included, A11, as it is the
only remaining CO detection by \citet{edge01} and \citet{salome03} 
not covered by our selection for a Spitzer observation.
This selection gives a sample of 61 line emitting BCGs to
which through an oversight in the compilation we added
one non-line emitting BCG from Crawford et al., A2622, to
give the total of 62 observed in this program and
listed in Table 1. The redshift distribution of the
sample is shown in Figure 1. At a redshift of $z=0.1$,
near the median of the sample, 1 arcsec corresponds to a distance of
1.84~kpc.

We believe the BCGs selected for this study provide a
direct test of the link between optical line emission and MIR emission
for individual objects but also for a statistically significant
and complete X-ray selected BCS sample.

\begin{figure*}
\plotone{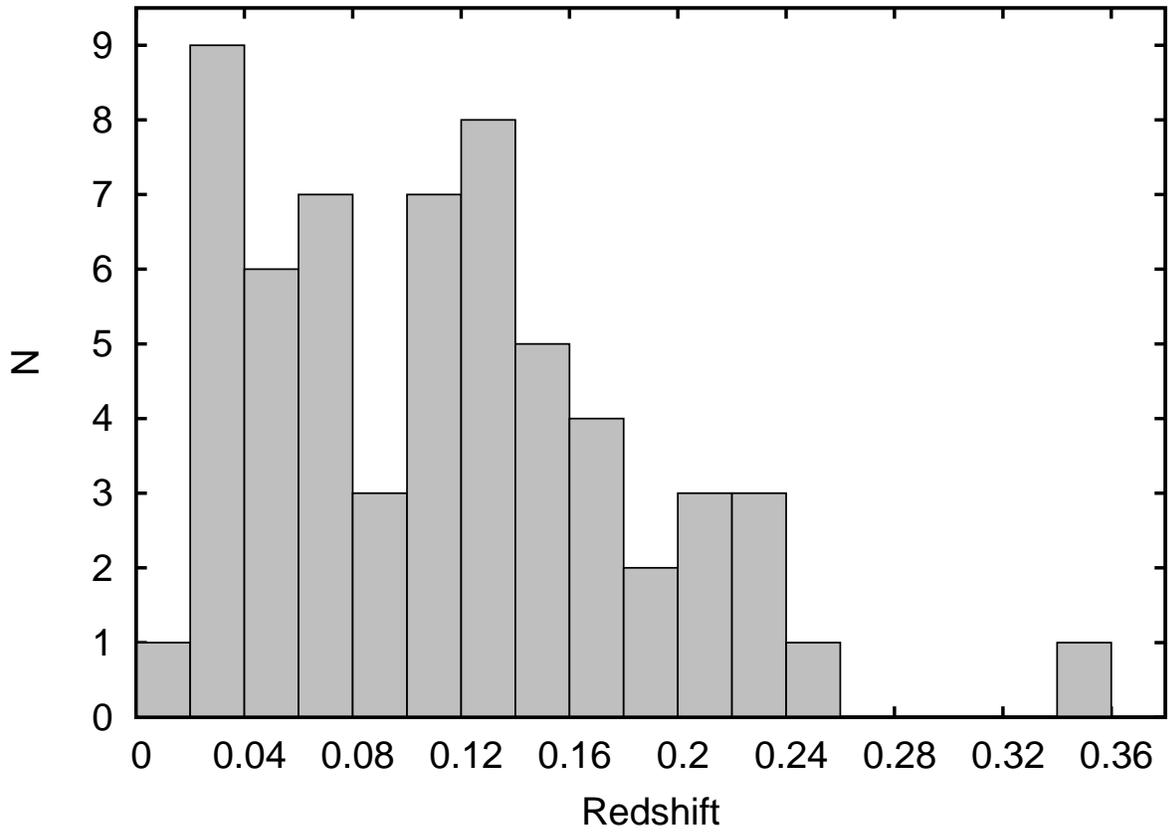}
\caption{Redshift histogram giving the number of clusters in our
sample in redshift bins.
\label{fig:43redshift}}
\end{figure*}

The 62 BCGs of our total observed 
sample are listed in Table \ref{tab:sample} and the redshift
distribution is shown in Figure \ref{fig:43redshift}. At a redshift of z=0.1, near the
median for the sample, 1 arcsec corresponds to a distance of 1.84 kpc. 


\subsection{IRAC data}

The 3.6, 4.5, 5.8 and 8.0 $\mu$m broad band images were obtained
during Cycle 3 (2007-2008)
using the Infrared Array Camera (IRAC; \citealt{fazio04})
on board the {\it Spitzer Space Telescope}.
One frame per band was observed with a 12 second frame times a 9 position
random medium step dither pattern resulting in 108
seconds total on source integration time.
Each image has a field of view of 5\farcm2 $\times$5\farcm2. 
Post basic calibrated data were used to make color maps
and generate photometric measurements.

Following the previous work by \citet{egami06}, we have begun our
study by measuring
fluxes in an aperture of diameter 12\farcs2.
Sky annulus radii and aperture corrections are listed
in the notes to Table \ref{tab:photo}.
The aperture photometry (as described here) is sufficiently accurate
to separate between AGNs and star forming objects and estimate
infrared luminosities, allowing a survey of the energetics of our sample
of BCGs.  We note that spectral energy distributions
created from our aperture photometry are not accurate enough to allow stellar
population models to be compared in detail.
Moreover, since our sample spans a range in distance and the IRAC images
exhibit color gradients, model spectral energy distributions should 
in future be fit to measurements from different regions in each galaxy.
We plan this in future investigations.

\subsection{MIPS data}

The 24.4 and 70 $\mu$m images were obtained using the
Multiple Imaging Photometer for Spitzer (MIPS; \citealt{rieke04}).
The exposure times at 24~$\mu$m and 70~$\mu$m were 3 cycles 
of 10 seconds each.
Basic calibrated data were reduced with the
MOPEX \citep{makovoz05} software which
performs interpolation and co-addition of FITS images.
The final 24 and  70~$\mu$m images have a field of
view of 7\farcm5$\times$8\farcm2 and 3\arcmin$\times$8\arcmin, respectively. 
To make photometric measurements we have used the same
aperture radii and corrections as
\citet{egami06}; these are listed in the notes to Table \ref{tab:photo}.

\subsection{Comparison data}

Comparison data in other spectral bands for the BCGs in our sample are listed
in Table \ref{tab:sample}.
Brightest cluster galaxies could host star formation, an active
galactic nucleus, as well as a cooling flow.  
When available, Table \ref{tab:sample} lists
X-ray, radio (1.4 GHz) and H$\alpha$ luminosities
and $[$OIII$]$(5007)/H$_\beta$ flux ratios.
X-ray luminosities provide a constraint on the mass in and radiative losses
from the hot ICM.  The H$\alpha$ recombination
line is excited by emission from newly formed hot stars 
or from an AGN.  
To discriminate between the presence of an
AGN and star formation we have sought a measure 
of the radiation
field in the $[$OIII$]$(5007\AA)/H$\beta$ optical line ratio
and an estimate of the radio luminosity.  

X-ray luminosities in almost all cases, are based on 0.1-2.4~keV ROSAT fluxes.
H$\alpha$ luminosities are based on measurements by \citet{crawford99},
or in a few cases drawn or estimated from the literature (see
citations given in the notes for Table \ref{tab:sample}).
The $[$OIII$]$(5007) to H$\beta$ flux ratios were
taken from measurements in Table 6 by \citet{crawford99} and from
spectroscopic line measurements listed in the Sloan Digital Sky Survey (SDSS)
archive.
Comparison VLA 1.4 GHz radio fluxes were drawn from 
the Faint Images of the Radio Sky at Twenty cm (FIRST) survey
\citep{becker95,white97}
and the NRAO VLA Sky Survey (NVSS)
\citep{condon98}. These have spatial resolutions
of 5\arcsec (B configuration) and 45\arcsec (D configuration), and
noise levels of $\sim 140$ and 450~$\mu$Jy, respectively.
Luminosities at 1.4~GHz were computed using integrated
flux densities\footnote{NVSS: http://www.cv.nrao.edu/nvss/NVSSlist.shtml
\\ FIRST: http://sundog.stsci.edu/cgi-bin/searchfirst}.
For a few objects, multiple sources were within a few
arcseconds of the BCG galaxy position.   In these cases
we looked at the images.   When objects appeared to be radio
doubles we used the NVSS integrated flux (see notes listed in Table 1).

\section{Results}

We discuss the morphology of the objects and the spectral energy
distributions which show that about 1/3 of the BCGs exhibit 
some form of IR excess.

\subsection{Morphology }

In Figure \ref{fig:images} 
we show small (2\arcmin$\times$2\arcmin) 
gray scale images of all the galaxies in our sample.
A number of them show bright sources at 24 and 70~$\mu$m.
All but one of the sources (Z9077)
was detected at 24~$\mu$m, but only about 1/4 of the sample
were detected at 70~$\mu$m.
Also shown in Figure \ref{fig:images}  are color maps made from sky
subtracted 8.0 and 3.6~$\mu$m
IRAC images.
Abell 1068, Abell 2146 , R0821+08 and Zw~2089 have prominent unresolved red
central sources.
NGC 4104 and Abell 11 have faint red and unresolved red nuclei.
The remaining BCGs have resolved weak (e.g., Abell 1664) or no color gradients
evident in the color maps.

Two of the objects in our survey (R0751+50 and Abell 2627)
are binary; each has two potential BCGs.
For R0751+50 the eastern galaxy contains a nucleus that is quite bright
at 24~$\mu$m but has no strong color gradient
in the 3.6 to 8.0 $\mu$m ratio color map, and so is probably forming stars. 
For Abell 2627 the northern
galaxy is bright at 24~$\mu$m and is also likely to be forming stars.

The galaxy cluster with the most interesting morphology is R0751+50 which contains
a bright oval or elliptical ring at seen most prominently at 8~$\mu$m
surrounding the two elliptical binary galaxies. The ring
may be a large spiral arm, is seen primarily at 8~$\mu$m but is also
faintly seen at 24~$\mu$m.  At a redshift of 0.0236 the
ring has a radius from the midpoint between the two
ellipticals of 30\arcsec $\sim 14$~kpc.  A faint counterpart is seen in 
Galaxy Evolution Explorer ({\it GALEX}) images
suggesting that the ring is also emitting in the UV.  The bright 8~$\mu$m
emission in the ring may be due to a dust emission feature at 7.7~$\mu$m.


\subsection{Spectral energy distributions}

In Figure~\ref{fig:sed}  we show spectral energy distributions 
for all of the BCGs using
aperture photometry listed in Table \ref{tab:photo}.
To minimize the contribution from the galaxy stellar component
we have used apertures for the IRAC measurements that are smaller
than those for the MIPS photometric measurements.
Small apertures cannot be used for long wavelength photometric
measurements because of the larger diffraction pattern; however,
the mid-infrared excess and longer wavelength emission are
unresolved in these larger apertures. Thus
the spectral energy distributions give
us an estimate for the contribution at all plotted
wavelengths within the IRAC apertures.

We have classified the 
spectral energy distributions into 4 different types 
depending on the flux ratio $F_{8.0 \mu {\rm m}}/F_{5.8 \mu{\rm m}}$.
We chose this flux ratio as
it increases when there is emission
from the PAH feature at 7.7~$\mu$m prominent in
star forming galaxies.
This flux ratio is also more robustly measured compared to
the $F_{24 \mu{\rm m }}/F_{8 \mu{\rm m}}$ color ratio which contains
much more scatter, although the two colors are correlated in our sample.
The correlation can be seen clearly in the slopes 
of the spectral energy distributions;
sources that have red 8.0 to 5.8~$\mu$m flux ratios tend to have red 24 to 
8~$\mu$m flux ratios.
This correlation can also be seen in Figure \ref{fig:colorcolor}
from the 5.8/3.6~$\mu$m  and 24/8~$\mu$m  colors.

In Figures \ref{fig:sed}a,b we show
the 13 sources that have the reddest color ratios;
$F_{8 \mu {\rm m}}/F_{5.8 \mu{\rm m}} > 1.3$.
These sources have clear mid-infrared excesses
(compared to a quiescent elliptical
galaxy stellar population), and are
bright at 24~$\mu$m and 70~$\mu$m.  All of this color group were detected
in both MIPS bands.  They are separated into two figures to highlight
those three with extreme red color (in Figure \ref{fig:sed}a)
throughout all bands and those with red colors primarily
in the 8.0/5.8~$\mu$m flux ratio but not the 4.5/3.6~$\mu$m flux ratio
(in Figure \ref{fig:sed}b).
The objects shown in Figure \ref{fig:sed}b could have elevated
emission at 8~$\mu$m due to a 7.7~$\mu$m PAH emission feature that
is prominent in star forming galaxies but are lacking hot dust
associated with an AGN at shorter wavelengths.
The dust emission feature may account for the high 8~$\mu$m flux compared
to that at 24 microns for R0338+09.

Sources with $1.0< F_{8 \mu {\rm m}}/F_{5.8 \mu{\rm m}} < 1.3$
and $0.75< F_{8\mu {\rm m}}/F_{5.8\mu{\rm m}} < 1.0$ are shown
in Figure \ref{fig:sed}c, and Figure \ref{fig:sed}d,e, respectively.
The division in color space is made so that not too many spectral
energy distributions lie
on the same plot.  In Figures \ref{fig:sed}f,g,h we show sources with
color $F_{8\mu {\rm m}}/F_{5.8\mu{\rm m}} < 0.75$.
Most of the objects with $F_{8\mu {\rm m}}/F_{5.8\mu{\rm m}} < 1.0$
were not detected at 70$\mu$m. Those with detected emission at
70~$\mu$m are also likely to contain dust that is heated by star formation
even if they don't exhibit red colors at wavelengths below 8 or 24 microns.
We note that the 70~$\mu$m images contain spurious structure
in the sky background
making accurate photometry difficult.
To list an object as detected at 70~$\mu$m we required a clear and statistically
significant source to be located at the expected galaxy position.
The objects with
color $F_{8 \mu {\rm m}}/F_{5.8 \mu{\rm m}} < 0.75$ and lacking
a detection at 70~$\mu$m are
consistent with a quiescent stellar population dominated by
emission from old stars.

It is clear from Figures \ref{fig:sed}a-h that
a significant number of the BCGs in our sample have infrared excesses.
22 objects of 62 are detected at 70$\mu$m, 18 have
8 $\mu$m to 5.8 ~$\mu$m flux ratios above 1.0 and 28 objects
have 24 $\mu$m to 8 $\mu$m flux ratios about 1.0.
Taking these three criteria together 35 objects of 62 have
some kind of excess.
As discussed by \citet{odea07} the infrared excesses are likely due to star 
formation with the exception of the four objects that we discuss below
that may be dominated by a dusty AGN.

We can consider the fraction of objects with IR excess
that did not display H$\alpha$ emission.
Only one object in our sample was reported lacking emission lines,
Abell 2622 \citep{owen95}.  
We did not find any evidence of infrared excess from this galaxy.
7 objects BCGs in our sample exhibited emission in NII and not
H$\alpha$, of these 4 exhibited some form of infrared excess 
(A2055, A2627, R0000+08, and Z4905)
and 3 did not (A1767, A2033, A4059).
The fraction with some form of infrared excesses 
is not significantly different than
that of the total sample.
Unfortunately we lack a comparison sample of X-ray bright
BCGs that lack optical emission lines.
Correlations between infrared excess and other properties are 
discussed further in paper II.

\subsubsection{Four Luminous AGN}

The reddest sources (those in Figure \ref{fig:sed}a) are an interesting group
containing Abell 2146, Abell 1068, and Zwicky 2089.
These objects have red colors in all IRAC band ratios suggesting that
very hot dust (a temperatures greater than 1000~K),
emitting even at 4.5 microns is present (e.g., as seen for Seyfert galaxies,
\citealt{alonso03}).
The short wavelength spectral energy distributions for these 3 objects
differ from that of a quiescent population and from that
of the objects with infrared excesses studied by \citet{egami06}.
Abell 2146, Abell 1068 and Zwicky 2089 have
$[$OIII$]$(5007)/H$\beta$ ratios of 7.1, 3.7, and 6.2 respectively,
consistent with the presence of a hard radiation source or AGN.
The only other object in our sample with a high $[$OIII$]$/H$\beta$ flux ratio
in the \citealt{crawford99} spectral study is Zw~3179 with
$[$OIII$]$/H$\beta =4.8^{+2.2}_{-2.2}$ but both lines are weak
so this ratio is uncertain. Therefore, there appears to be
a strong correspondence between the optical line ratios and the presence
of hot dust emission in the IRAC bands.

Compared to the other objects with red mid-infrared colors,
$F_{8\mu {\rm m}}/F_{5.8\mu{\rm m}} > 1.3$,
Abell 2146 has a blue color in the
ratio $F_{70\mu {\rm m}}/ F_{24 \mu {\rm m}}$.  Abell 2146 has a ratio of
$\sim 3.7$
compared to $F_{70\mu {\rm m}}/ F_{24 \mu {\rm m}} \sim 7$  for Zw~2089
and 12 for Abell 1068.  Abell 2146 could have a warmer
dust temperature than the other objects with mid-infrared excesses suggesting
that it is AGN dominated \citep{sanders88,armus07}.
This BCG could have a spectral energy distribution similar to that
of a Seyfert galaxy or a quasar.
Abell 2146 has previously been classified as an AGN \citep{allen95}, 
whereas previous studies found 
that Abell 1068 hosts an extended starburst \citep{allen95,mcnamara04b}.

Four BCGs have prominent red unresolved nuclear sources as seen
from the 8.0/3.6$\mu$m  color maps: Abell 1068, Abell 2146, Zw 2089
and R0821+07.  The first three have evidence for hot dust.
R0821+07 has a red  color
$F_{8\mu {\rm m}}/F_{5.8\mu{\rm m}} > 1.3$ but no evidence
for hot dust from its
$F_{4.5\mu {\rm m}}/F_{3.6\mu{\rm m}}$ ratio.
It also has a high ratio
$[$OIII$]$/H$\beta \sim 2.5$ ratio, so it too could
host a more obscured dusty AGN.

\subsection{Estimated infrared luminosities}

Following the procedure used by \citet{egami06}
we use the 8 and 24$\mu$m fluxes to estimate the infrared luminosity.
Previous studies (e.g., \citealt{spinoglio,elbaz02}) have found that
bolometric luminosities estimated from
the 15~$\mu$m flux are less sensitive to dust temperature than
those estimated from fluxes at other wavelengths.  We estimate
the flux at 15~$\mu$m with a linear interpolation of the 8 and 24~$\mu$m fluxes.
Then the 15~$\mu$m flux is converted to an infrared luminosity $L_{\rm IR}$
using equation 13 by \citet{elbaz02} giving a relation
between 15~$\mu$m observations and total infrared luminosity
that has been established from {\it Infrared Space Observatory} 
(ISO) observations of quiescent and active galaxies;
\begin{equation}
L_{\rm IR} = 11.1^{+5.5}_{-3.7} \times (\nu L_\nu [15\mu {\rm m}])^{0.998}.
\label{eqn:LTIR}
\end{equation}
where both $L_{\rm IR}$ and $\nu L_\nu [15\mu {\rm m}]$ are in erg/s.
Infrared luminosities are listed in Table \ref{tab:selected}.
For galaxies that have no evidence of an infrared excess and
have not been detected at 70~$\mu$m, this procedure 
significantly overestimates the infrared luminosity.
We have checked that this estimate is approximately consistent
with the luminosity at 70$\mu$m for the objects
that are not red at wavelengths shorter than 8~$\mu$m
but are detected at 70~$\mu$m 
(e.g., that are shown in Figure \ref{fig:sed}h).
Consequently we have listed
infrared luminosities only for objects that have evidence
of an infrared excess.
We list only objects with $F_{8.0}/F_{5.8\$mu {\rm m}} > 1.0$, or
$F_{24}/F_{8\mu {\rm m}} > 1.0$, 
or that were detected at 70$\mu$m.
We find that 9 of the BCGs can be classified as LIRGs with
$10^{11} L_\odot < L_{\rm IR} < 10^{12} L_\odot$.
The 9  LIRGs include the 4 that host 
dusty AGN. These 4 have luminosities which put them
in the Seyfert/QSO transition region for IR luminosity (e.g., 
\citealt{rush93,schweitzer06}).  Thus, these are Type 2 luminous AGN. 

\section{Summary}

We have presented Spitzer IRAC and MIPS photometry of a sample of 62 BCGs 
selected on cluster X-ray flux and BCG H$\alpha$ or [NII] flux 
to favor objects in cool cluster cores. 
We find that one half to one-third of the sample show an infrared excess
above that expected for the old stellar population. These
results confirm the previous results of \citet{egami06},
based on a sample of 11 BCGs in X-ray luminous clusters, that some
BCGs in cool core clusters show an infrared excess due to star formation. 

An interesting difference between our survey and previous studies
\citep{egami06,donahue07}
is the discovery of a small number of objects
that host dusty AGNs,
thanks to our much larger sample.
Three galaxies
exhibit red 4.5/3.6$\mu$m flux ratios and unresolved
nuclei seen in IRAC color maps indicating the presence
of hot dust.
These three, Abell 1068, Abell 2146, and Zwicky 2089, have measured 
high $[$OIII$]$(5007)/H$\beta$ flux ratios
suggesting that they host a dusty AGN.
An additional BCG, R0821+07, has
a red unresolved nucleus at 8$\mu$m and a high
$[$OIII$]$(5007)/H$\beta$ ratio suggesting that it too hosts a dusty AGN.

These four BCGs have infrared luminosities greater than
$10^{11} L_\odot$ and so can be classified as LIRGs.
A comparison between their infrared luminosity and
1.4 GHz radio luminosity yields ratios of $10^6 - 10^7$ suggesting
that they can be considered radio quiet.
The only similar previously known object is the
more distant ($z=0.44$) and hyperluminous infrared galaxy
IRAS 09104+4109 \citep{fabian95},
also a BCG in an X-ray luminous cluster.

In addition to the 4 hosting AGNs, there are 5 other BCGs with 
infrared luminosities greater than $10^{11} L_\odot$ 
that can be classified as luminous infrared galaxies (LIRGs).
Excluding the AGNs, the remaining brightest
cluster galaxies with infrared excesses
are likely bright in the mid-infrared
because of star formation as discussed in our companion paper \citep{odea07}.
In this work we have discussed the identifications and 
broad properties of this sample.
Our companion paper (paper II; \citealt{odea07}) 
searches for correlations between star formation rates,
radio, H$\alpha$, CO and X-ray luminosities and mass deposition rates estimated
from the X-ray observations.
Planned future work includes fitting model spectral energy distributions to
the observations.  

\acknowledgments
We thank Eiichi Egami for helpful correspondence.
This work is based on observations made with the Spitzer Space Telescope, which is operated 
by the Jet Propulsion Laboratory, California Institute of Technology under a contract with NASA. 
Support for this work at University of Rochester and Rochester Institute of Technology
was also provided by NASA through an award issued by JPL/Caltech and
by NSF grants AST-0406823 \& PHY-0552695.

{}
\clearpage

\begin{deluxetable}{llcrrrrrrrr}
\tabletypesize{\scriptsize}
\setlength{\tabcolsep}{0.02in} 
\rotate
\tablecaption{The brightest cluster galaxy sample
and comparison luminosities \label{tab:sample}}
\tablehead{
\colhead{Cluster} &
\colhead{RA} &
\colhead{DEC} &
\colhead{z} &
\colhead{$L_X$} &
\colhead{Ref.} &
\colhead{$L_{{\rm H}\alpha}$}  &
\colhead{$[$OIII$]$/H$\beta$}  &
\colhead{Ref.} &
\colhead{$L_{1.4{\rm GHz}}$}  &
\colhead{Ref.}  \\
\colhead{} &
\colhead{(2000)} &
\colhead{(2000)} &
\colhead{} &
\colhead{($10^{44}$erg~s$^{-1}$)}&
\colhead{} &
\colhead{($10^{40}$erg~s$^{-1}$)}&
\colhead{} &
\colhead{} &
\colhead{($10^{31}$erg~s$^{-1}$Hz$^{-1}$)}&
\colhead{}
}
\startdata

Abell 11      & 00 12 33.8 &$-$16 28 06 & 0.1510 & 0.99  & 1 & 120   &\nodata&3 & 6.0   & 1 \\

Abell 85      & 00 41 50.4 &$-$09 18 14 & 0.0551 & 5.55  & 2 & 0.4   &\nodata&1 & 0.42  & 1 \\

Abell 115     & 00 55 50.6 &$+$26 24 39 & 0.1970 &10.4   & 3 & 13.9  & 0.4   &2 & 158   & 1 \\

Abell 262     & 01 52 46.5 &$+$36 09 08 & 0.0166 & 0.31  & 3 & 0.3   & 1.1   &2 & 0.042 & 1 \\

Abell 291     & 02 01 43.1 &$-$02 11 47 & 0.1960 & 4.81  & 2 & 28.6  & 0.3   &2 & 1.4   & 2 \\

Abell 646     & 08 22 09.6 &$+$47 05 54 & 0.1268 & 2.76  & 3 & 17.3  & 0.8   &1 & 2.3   & 2 \\

Abell 795     & 09 24 05.3 &$+$14 10 22 & 0.1355 & 3.70  & 3 & 11.6  & 1.8   &2 & 5.5   & 2 \\

Abell 1068    & 10 40 44.4 &$+$39 57 12 & 0.1386 & 5.00  & 3 & 143   & 3.7   &1 & 0.46  & 2 \\

Abell 1084    & 10 44 32.9 &$-$07 04 08 & 0.1329 & 4.56  & 2 & 4.8   &\nodata&2 & 1.6   & 2 \\

Abell 1204    & 11 13 20.3 &$+$17 35 41 & 0.1706 & 4.96  & 3 & 6.7   & 1     &2 & 0.21  & 2 \\

Abell 1361    & 11 43 39.5 &$+$46 21 22 & 0.1167 & 2.27  & 3 & 8.1   & 0.9   &2 & 32    & 2 \\

Abell 1664    & 13 03 42.5 &$-$24 14 41 & 0.1276 & 2.60  & 2 & 67.2  & 0.5   &2 & 1.6   & 1 \\

Abell 1668    & 13 03 46.6 &$+$19 16 18 & 0.0640 & 0.95  & 3 & 1.3   & 1.7   &2 & 0.95  & 2 \\

Abell 1767    & 13 36 08.1 &$+$59 12 24 & 0.0715 & 1.51  & 3 &\nodata&\nodata&  & 0.023 & 2 \\

Abell 1885    & 14 13 43.6 &$+$43 39 45 & 0.0900 & 1.49  & 3 & 3.1   & 0.3   &2 & 0.92  & 2 \\

Abell 1930    & 14 32 37.9 &$+$31 38 49 & 0.1316 & 2.66  & 3 & 1.4   &\nodata&2 & 0.38  & 2 \\

Abell 1991    & 14 54 31.4 &$+$18 38 34 & 0.0595 & 0.86  & 3 & 0.7   &\nodata&2 & 0.33  & 2 \\

Abell 2009    & 15 00 19.6 &$+$21 22 11 & 0.1532 & 6.09  & 3 & 7.3   & 0.7   &2 & 5.3   & 1 \\

Abell 2033    & 15 11 26.6 &$+$06 20 58 & 0.0780 & 1.40  & 3 &\nodata&\nodata&  & 8.6   & 2 \\

Abell 2052    & 15 16 44.6 &$+$07 01 18 & 0.0351 & 1.39  & 3 & 1.6   & 4.2   &2 & 16    & 1 \\

Abell 2055    & 15 18 45.8 &$+$06 13 57 & 0.1019 & 2.98  & 3 &\nodata&\nodata&  & 14    & 1 \\

Abell 2072    & 15 25 48.7 &$+$18 14 11 & 0.1270 & 2.02$^a$&3& 3.1   & 0.7   &2 & 0.19  & 2 \\

Abell 2146    & 15 56 13.8 &$+$66 20 55 & 0.2343 & 6.67  & 4 & 92.4  & 7.1   &2 & 2.6   & 1 \\

Abell 2204    & 16 32 46.9 &$+$05 34 33 & 0.1514 & 14.1  & 3 & 119   & 0.7   &2 & 3.7   & 2 \\

Abell 2292    & 17 57 06.7 &$+$53 51 38 & 0.1190 & 1.26  & 5 & 2.7   & 0.3   &2 & 2.3   & 1 \\

Abell 2495    & 22 50 19.6 &$+$10 54 13 & 0.0808 & 1.97  & 3 & 1.5   &\nodata&2 & 0.25  & 1 \\

Abell 2622    & 23 35 01.4 &$+$27 22 22 & 0.0610 & 0.62  & 3 &\nodata&\nodata&  & 0.72  & 1 \\

Abell 2626    & 23 36 30.7 &$+$21 08 49 & 0.0552 & 1.07  & 3 & 0.53  &\nodata&2 & 0.41  & 1 \\

Abell 2627    & 23 36 42.4 &$+$23 55 06 & 0.1270 & 2.20  & 3 &\nodata&\nodata&  &$<$0.062& 2 \\

Abell 2665    & 23 50 50.6 &$+$06 09 00 & 0.0567 & 1.12  & 3 & 0.32  & 0.2   &2 & 0.44  & 1 \\

Abell 3112    & 03 17 57.7 &$-$44 14 18 & 0.0761 & 4.31  & 2 &\nodata&\nodata&  &\nodata& \nodata \\

Abell 4059    & 23 57 00.7 &$-$34 45 32 & 0.0475 & 1.80  & 2 &\nodata&\nodata&  & 7.1   & 1 \\

IC 1262       & 17 33 02.1 &$+$43 45 35 & 0.0331 & 0.30  & 3 & 0.10  & 1.5   &2 & 0.26  & 1  \\

NGC 4104      & 12 06 38.8 &$+$28 10 27 & 0.0281 & 0.11  & 3 & 0.33  &\nodata&2 & 0.004 & 2 \\

NGC 4325      & 12 23 06.6 &$+$10 37 17 & 0.0259 & 0.11  & 3 & 0.39  & 0.9   &1 & $<$0.001& 2 \\

NGC 5096      & 13 20 14.6 &$+$33 08 39 & 0.0360 & 0.14  & 3 & 0.31  & 2     &2 & 0.26  & 2 \\

NGC 6338      & 17 15 22.6 &$+$57 24 43 & 0.0282 & 0.24  & 3 & 0.66  & 0.8   &2 & 0.09  & 2 \\


PKS0745-191   & 07 47 31.35&$-$19 17 39.7& 0.1028& 16.4  & 6 & 70    &\nodata&3 & 66    & 1 \\

RXC J0000.1+0816& 00 00 07.1&$+$08 16 49 & 0.0400 & 0.22 & 3 &\nodata&\nodata&  & 0.32  & 1 \\

RXC J0338.6+0958& 03 38 40.5&$+$09 58 12 & 0.0338 & 2.20 & 3 & 9.6   & 0.6   &2 & 0.17  & 1 \\

RXC J0352.9+1941& 03 52 58.9&$+$19 41 00 & 0.1090 & 2.50 & 3 & 33.8  & 0.6   &2 & 0.57  & 1 \\

RXC J0439.0+0520& 04 39 02.2&$+$05 20 45 & 0.2080 & 6.28 & 3 & 70.6  & 1.2   &2 & 10.7  & 2 \\

RXC J0751.3+5012& 07 51 19.9&$+$50 14 07 & 0.0236 & 0.13 & 3 & 0.87  &\nodata&1 &\nodata&\nodata \\

RXC J0821.0+0751& 08 21 02.4&$+$07 51 47 & 0.1100 & 1.29 & 4 & 17.7  & 2.5   &2 & 0.066 & 2 \\

RXC J1442.3+2218& 14 42 19.4&$+$22 18 13 & 0.0970 & 1.63 & 3 & 2.7   & 1.7   &2 & 0.47  & 2 \\

RXC J1532.9+3021& 15 32 53.8&$+$30 21 00 & 0.3615 & 23.3 & 3 & 427   & 0.6   &1 & 22.9  & 1 \\

RXC J1720.1+2637& 17 20 10.1&$+$26 37 32 & 0.1611 & 10.6 & 3 & 25.4  & 0.2   &1 &  6.5  & 1 \\

RXC J2129.6+0005& 21 29 39.9&$+$00 05 23 & 0.2346 & 13.9 & 3 & 28.5  & 0.2   &1 & 4.1   & 2 \\

S1101          & 23 13 58.8 &$-$42 43 38 & 0.0564 & 1.84 & 2 &9.98   &\nodata&4 &\nodata&\nodata \\

Zw235          & 00 43 52.1 &$+$24 24 22 & 0.0830 & 1.93 & 3 & 2.3   & 0.8   &2 & 0.87  & 1 \\

Zw348          & 01 06 49.3 &$+$01 03 23 & 0.2535 & 7.37 & 4 & 195   & 0.6   &1 & 0.71  & 2 \\

Zw808          & 03 01 38.2 &$+$01 55 15 & 0.1690 & 4.20 & 3 & 3.3   &\nodata&2 & 32.3  & 1 \\

Zw1665         & 08 23 21.7 &$+$04 22 22 & 0.0311 & 0.22 & 3 & 0.41  &\nodata&1 & 0.033 & 2 \\

Zw2089         & 09 00 36.8 &$+$20 53 43 & 0.2350 & 8.10 & 3 & 265   & 6.2   &  & 1.5   & 2 \\

Zw2701         & 09 52 49.2 &$+$51 53 06 & 0.2150 & 7.89 & 3 & 4.6   & 1.3   &1 & 1.9   & 2 \\

Zw3179         & 10 25 58.0 &$+$12 41 09 & 0.1432 & 3.14 & 3 & 6.6   & 4.8   &1 & 5.1   & 2 \\

Zw3916         & 11 14 21.9 &$+$58 23 20 & 0.2040 & 4.51 & 4 & 10.6  & 0.8   &1 & 8.2   & 1 \\

Zw4905         & 12 10 16.8 &$+$05 23 11 & 0.0766 & 0.70 & 3 &\nodata&\nodata&  &$<$0.006& 2 \\
Zw8193         & 17 17 19.1 &$+$42 26 59 & 0.1754 & 8.04 & 4 & 95.6  & 0.8   &2 & 11.8  & 2 \\
Zw8197         & 17 18 11.8 &$+$56 39 56 & 0.1140 & 1.85 & 3 & 8.5   & 1.0   &1 & 1.2   & 2 \\
Zw8276         & 17 44 14.5 &$+$32 59 30 & 0.0750 & 2.34 & 3 & 9.7   & 0.8   &2 & 1.3   & 1 \\
Zw9077         & 23 50 35.4 &$+$29 29 40 & 0.150  & 3.48 & 3 & 16.4  &\nodata&2 & 4.2   & 1
\enddata
\tablecomments{
Columns 2 and 3 list the position of the brightest cluster galaxies.
Most of these positions are those listed by \citet{crawford99}.
Column 4 lists redshifts.
Column 5 lists X-ray luminosities based on
0.1-2.4~keV ROSAT fluxes, excepting for Abell 11 that is 0.5-2~keV.
These have been corrected to be consistent with a Hubble constant
$H_0 = 70$~Mpc$^{-1}$~km~s$^{-1}$.
These luminosities have been taken from measurements listed
in column 6.
References --
1. \citet{david99}; 
2. \citet{bohringer04}; 
3. \citet{ebeling98}; 
4. \citet{ebeling00}; 
5. \citet{brinkmann95}; 
6. \citet{edge90}. 
Column 7 lists luminosities in H$\alpha$.
Column 8 lists $[$OIII$]$(5007)/H$\beta$ ratios.
Optical line fluxes and ratios are derived from measurements
described in column 9.
References ---
1. SDSS Spectral line measurements.
2. Tables 5 and 6 by \citet{crawford99}.
3. Based on Pa$\alpha$ fluxes measured by \citep{edge02}
4. Table 1 by \citet{jaffe05}.
and an intrinsic H$\alpha$/Pa$\alpha$ ratio of
8.46 for case B recombination at a temperature of 10$^4$~K and
a density of 100~cm$^{-3}$.
Column 10 lists integrated luminosities per Hz at 1.4~GHz.
The radio fluxes are taken from the surveys
listed in column 11.
The objects lacking luminosities were
not at positions covered by either NVSS or FIRST surveys.
References --
1. NVSS survey \citep{condon98};
2. FIRST survey \citep{becker95,white97};
Notes on individual radio observations:
The following sources were extended and so the NVSS flux
was used, not the FIRST flux:  A85, A2052, A2055, R1720+26, and Z3916.
The following had multiple components seen in the FIRST survey and
these were added to estimate the total:  A1361, A2033, and NGC5096.
The following were extended in the NVSS images so fits images were
used to integrate the total:  A2009 and IC 1262, R0338+09, R1532+30.
The non detections were given 3-sigma upper limits.
Notes: $^a$.  The Rosat flux is contaminated by an AGN so the cluster X-ray 
luminosity is less than the number given here.
Additional notes:  We now describe objects lacking H$\alpha$ luminosities
in the table.
7 objects are classified as showing 
optical emission lines but lacking H$\alpha$ emission.   
Objects with [NII] emission
but no H$\alpha$ reported by \citet{crawford99} are
are Abell 1767, Abell 2033, Abell 2055,  Abell 2627, 
R0000+0816, and Z4905.
Abell 4059 exhibits a similar spectrum with [NII] but
no H$\alpha$ as seen
from the 6dFGS archive \citep{jones04,jones06}.
A3112 exhibits [OII] emission \citep{katgert98} but we have
failed to find an report of its H$\alpha$ emission.
Through an over sight A2622 was included in the sample
as it lacks emission lines \citep{owen95}.
}
\end{deluxetable}
\clearpage

\begin{deluxetable}{lrrrrrrrr}
\tablecaption{Infrared Photometry
\label{tab:photo}}
\tablehead{
\colhead{Cluster} &
\colhead{3.6$\mu$m} &
\colhead{4.5$\mu$m} &
\colhead{5.8$\mu$m} &
\colhead{8$\mu$m} &
\colhead{24$\mu$m} &
\colhead{70$\mu$m} \\
\colhead{} &
\colhead{(mJy)}&
\colhead{(mJy)}&
\colhead{(mJy)}&
\colhead{(mJy)}&
\colhead{(mJy)}&
\colhead{(mJy)}&
\colhead{}
}
\startdata
      A0011 &     1.36 &     1.03 &     0.89 &     1.71 &    10.59 &       82  \\

      A0085 &     5.57 &     3.45 &     2.46 &     1.58 &     1.88 & $<$   20  \\

      A0115 &     1.06 &     0.80 &     0.45 &     0.51 &     0.50 & $<$   20  \\

      A0262 &    16.93 &     9.97 &     7.96 &     6.57 &     4.03 &       88  \\

      A0291 &     1.16 &     0.87 &     0.54 &     0.50 &     0.53 & $<$   20  \\

      A0646 &     1.78 &     1.27 &     0.85 &     0.77 &     2.52 & $<$   20  \\

      A0795 &     1.96 &     1.36 &     0.91 &     0.68 &     0.25 & $<$   20  \\

      A1068 &     3.41 &     2.98 &     3.66 &     9.23 &    74.47 &      941  \\

      A1084 &     1.52 &     1.08 &     0.70 &     0.44 &     0.31 & $<$   20  \\

      A1204 &     1.23 &     0.94 &     0.59 &     0.55 &     1.41 & $<$   20  \\

      A1361 &     2.42 &     1.68 &     1.10 &     0.81 &     0.66 & $<$   20  \\

      A1664 &     2.27 &     1.61 &     1.34 &     2.67 &     4.01 &       78  \\

      A1668 &     4.61 &     2.90 &     2.07 &     1.41 &     0.73 & $<$   20  \\

      A1767 &     5.89 &     3.81 &     2.61 &     2.03 &     1.03 & $<$   20  \\

      A1885 &     1.68 &     1.24 &     1.04 &     1.02 &     3.81 & $<$   20  \\

      A1930 &     2.21 &     1.56 &     0.93 &     0.68 &     0.76 & $<$   20  \\

      A1991 &     4.53 &     2.84 &     1.91 &     1.35 &     0.98 & $<$   20  \\

      A2009 &     1.84 &     1.34 &     0.82 &     0.68 &     0.66 & $<$   20  \\

      A2033 &     5.21 &     3.37 &     2.23 &     1.45 &     0.86 & $<$   20  \\

      A2052 &     7.34 &     4.50 &     3.53 &     2.76 &     4.96 &       95  \\

      A2055 &     3.37 &     2.67 &     2.26 &     2.09 &     2.82 & $<$   20  \\

      A2072 &     2.28 &     1.48 &     0.87 &     0.75 &     0.73 & $<$   20  \\

      A2146 &     1.18 &     1.19 &     1.37 &     2.53 &    24.10 &       90  \\

      A2204 &     3.31 &     2.43 &     1.67 &     1.79 &     2.85 &       34  \\

      A2292 &     3.08 &     2.11 &     1.27 &     0.90 &     1.00 &       25  \\

      A2495 &     2.99 &     1.91 &     1.27 &     0.86 &     0.84 & $<$   20  \\

      A2622 &     5.64 &     3.53 &     2.48 &     1.59 &     1.11 & $<$   20  \\

      A2626 &     5.50 &     3.47 &     2.40 &     1.71 &     1.39 & $<$   20  \\

      A2627 &     3.43 &     2.75 &     2.22 &     2.23 &     0.86 & $<$   20  \\

      A2665 &     6.57 &     4.02 &     2.86 &     1.97 &     0.99 & $<$   20  \\

      A3112 &     5.15 &     3.43 &     2.46 &     2.16 &     3.09 & $<$   20  \\

      A4059 &     7.91 &     4.86 &     3.47 &     2.68 &     1.91 & $<$   20  \\

     IC1262 &     7.59 &     4.54 &     3.27 &     2.26 &     1.21 & $<$   20  \\

    NGC4104 &    17.51 &    10.60 &     9.45 &    12.38 &    27.90 &      540  \\

    NGC4325 &    10.04 &     5.96 &     4.46 &     3.41 &     2.36 & $<$   20  \\

    NGC5096 &     7.15 &     4.33 &     3.09 &     2.02 &     1.32 & $<$   20  \\

    NGC6338 &    17.77 &    10.45 &     7.73 &     5.11 &     3.16 &       30  \\

PKS0745-191 &     3.95 &     2.76 &     2.12 &     2.87 &    10.34 &      154  \\

   R0000+08 &     6.95 &     4.20 &     3.07 &     2.22 &     2.42 &       53  \\

   R0338+09 &     9.43 &     5.70 &     4.55 &     9.61 &     2.39 &       78  \\

   R0352+19 &     2.04 &     1.57 &     1.41 &     2.07 &     5.14 &      132  \\

   R0439+05 &     1.71 &     1.36 &     0.85 &     0.92 &     2.10 &       55  \\

   R0751+50 &    13.14 &     7.80 &     5.87 &     3.94 &     2.58 &       44  \\

   R0751+50 &    13.62 &     8.09 &     6.02 &     3.78 &     1.81 & $<$   20  \\

   R0821+07 &     2.52 &     1.81 &     1.62 &     7.03 &    18.00 &      327  \\

   R1442+22 &     2.83 &     1.89 &     1.26 &     0.97 &     0.63 & $<$   20  \\

   R1532+30 &     0.82 &     0.71 &     0.48 &     0.95 &     3.77 &      112  \\

   R1720+26 &     2.11 &     1.55 &     0.93 &     0.81 &     0.48 & $<$   20  \\

   R2129+00 &     1.45 &     1.16 &     0.74 &     0.56 &     1.04 & $<$   20  \\

      S1101 &     4.90 &     3.05 &     2.15 &     1.60 &     1.27 & $<$   20  \\

      Z0235 &     3.30 &     2.14 &     1.39 &     0.90 &     0.48 & $<$   20  \\

      Z0348 &     0.63 &     0.58 &     0.43 &     1.13 &     4.54 &       44  \\

      Z0808 &     1.95 &     1.44 &     0.85 &     0.58 &     0.44 & $<$   20  \\

      Z1665 &     8.20 &     4.88 &     3.63 &     2.31 &     1.59 & $<$   20  \\

      Z2089 &     0.94 &     1.15 &     1.77 &     3.94 &    33.59 &      235  \\

      Z2701 &     1.21 &     0.92 &     0.57 &     0.37 &     0.36 & $<$   20  \\

      Z3179 &     2.97 &     2.12 &     1.28 &     0.91 &     0.35 & $<$   20  \\

      Z3916 &     1.43 &     1.12 &     0.62 &     0.49 &     0.41 & $<$   20  \\

      Z4905 &     3.54 &     2.26 &     1.47 &     1.02 &     1.09 & $<$   20  \\

      Z8193 &     4.92 &     3.73 &     2.40 &     3.99 &    10.66 &      177  \\

      Z8197 &     2.17 &     1.48 &     0.97 &     0.99 &     0.85 & $<$   20  \\

      Z8276 &     3.24 &     2.15 &     1.63 &     1.62 &     3.22 &       22  \\

      Z9077 &     1.58 &     1.04 &     0.70 &     0.44 & $<$  0.4 & $<$   20

\enddata
\tablecomments{
Aperture photometry at 3.6, 4.5, 5.8, 8 and 24.4 $\mu m$ were measured
from the Band 1-4 IRAC images and Band 1 MIPS images.
For the IRAC images we
used an aperture diameter of 12\farcs2
and a sky annulus with inner and outer radii of
32\arcsec and 48\arcsec.   Aperture corrections taken from Table 5.7 of the
IRAC data handbook were applied.   These were 1.05, 1.05, 1.06,
and 1.07 for Bands 1-4, respectively.
At 24$\mu$m we used an aperture diameter of 26\arcsec
and a sky annulus with inner and outer
radii of 20\arcsec and 32\arcsec as did \citet{egami06}.
Upper limits at $24\mu$m and $70\mu$m were estimated from
the fluxes of the faintest sources in the field.
At $70\mu$m  we used an aperture diameter of 35\arcsec
and a sky annulus with inner and outer
radii of 39\arcsec and 65\arcsec as did \citet{egami06}.
The aperture correction
for the 24$\mu$m and $70\mu$m fluxes is 1.167 and 1.308,
respectively based on Table 3.13 of
the MIPS data handbook and is the same as that used
by \citet{egami06}.
R0751+50 contains a pair of bright elliptical galaxies, hence
two measurements are given, one for each galaxy.
Abell 2627 is a galaxy pair; photometry is only given
for the northern galaxy.
R0338+09 has a nearby bright star which is adding to
the uncertainty of the photometry.
}
\end{deluxetable}
\vfill\eject
\clearpage

\begin{deluxetable}{lc}
\tablewidth{300pt}
\tablecaption{Estimated Infrared Luminosities of BCGs
\label{tab:selected}}
\tablehead{
\colhead{Cluster} &
\colhead{L$_{\rm IR}$} 
\\
\colhead{}&
\colhead{($10^{44}$erg s$^{-1}$)} 
}
\startdata
Z2089*	  & 64.68   \\
A2146*	  & 45.46   \\
A1068*	  & 44.61    \\ 
R0821+07* & 8.47    \\  \hline
R1532+30* & 22.62    \\
Z8193*	  & 13.70    \\
Z0348*	  & 11.92    \\
A0011*	  & 7.97     \\
PKS0745-1& 3.80     \\
A1664	 & 3.21     \\
R0352+19 & 2.40     \\
NGC4104	 & 0.80     \\
R0338+09 & 0.39     \\ \hline
R0439+05* & 4.17    \\
A2204    & 3.23     \\
A2627	 & 1.59     \\
A0115	 & 1.30     \\
Z8197	 & 0.72     \\ \hline
R2129+00 & 2.93     \\
A1204	 & 1.73     \\
A0646	 & 1.49     \\
A2055	 & 1.46     \\
A0291    & 1.30  \\
A1885	 & 1.04     \\
A3112	 & 0.84     \\
A2292	 & 0.80     \\
A1930    & 0.75  \\
Z8276	 & 0.74     \\
Z4905    & 0.29 \\
A0085    & 0.28  \\ 
A2052	 & 0.24     \\
R0000+08 & 0.20     \\
NGC6338	 & 0.18     \\
R0751+50 & 0.10     \\ 
A0262    & 0.08     \\
\hline
A0795    & $<$0.5  \\
A1084    & $<$4.1  \\
A1361    & $<$0.6  \\
A1668    & $<$0.3  \\
A1767    & $<$0.5  \\
A1991    & $<$0.3  \\
A2009    & $<$1.0  \\
A2033    & $<$0.4  \\
A2072    & $<$0.7  \\
A2495    & $<$0.3  \\
A2622    & $<$0.3  \\
A2626    & $<$0.3  \\
A2665    & $<$0.3  \\
A4059    & $<$0.3  \\
IC1262   & $<$0.1  \\
NGC4325  & $<$0.1  \\
NGC5096  & $<$0.1  \\
R1442+22 & $<$4.5  \\
R1720+26 & $<$1.1  \\
S1101    & $<$0.3  \\
Z235     & $<$0.3  \\
Z808     & $<$1.0  \\
Z1665    & $<$0.1  \\
Z2701    & $<$1.1  \\
Z3179    & $<$0.8  \\
Z3916    & $<$1.3  \\
Z9077    & $<$0.6
\enddata
\tablecomments{
Infrared luminosities are estimated from the flux at 15$\mu$m 
as mentioned in Section 3.3 for BCGs that are either detected
at 70$\mu$m, have color ratio
$F_{8 \mu {\rm m}}/F_{5.8 \mu {\rm m}}>1.0$ or
$F_{24 \mu {\rm m}}/F_{8.0 \mu {\rm m}}>1.0$.
The top section contains four BCGs that are suspected
to harbor dusty Type II AGNs. Z2089, A2146 and A1068 exhibit a
red $F_{4.5 \mu {\rm m}}/F_{3.6\mu {\rm m}} $ color and all four
have high $[$OIII$]$(5007)/H$\beta$ flux ratios.
The second set is
the remaining 10 BCGs with $F_{8 \mu {\rm m}}/F_{5.8 \mu {\rm m}}>1.3$.
The third section is the set of 6 clusters with
$1.0<F_{8 \mu {\rm m}}/F_{5.8 \mu {\rm m}}<1.3$.  
The fourth set is the remaining BCGs with IR excesses.
Specifically they have ratios
$F_{8 \mu {\rm m}}/F_{5.8 \mu {\rm m}}>1.0$,
$F_{24 \mu {\rm m}}/F_{8.0 \mu {\rm m}}>1.0$,
or a detected 70$\mu$m flux.
The BCGs marked with a $*$ can be classified as LIRGs
since they have L$_{IR}$ greater than $10^{11}L_\odot$.
The last section contains upper limits for
the remaining objects.
}
\end{deluxetable}
\vfill\eject
\clearpage

\begin{figure*}
\plotone{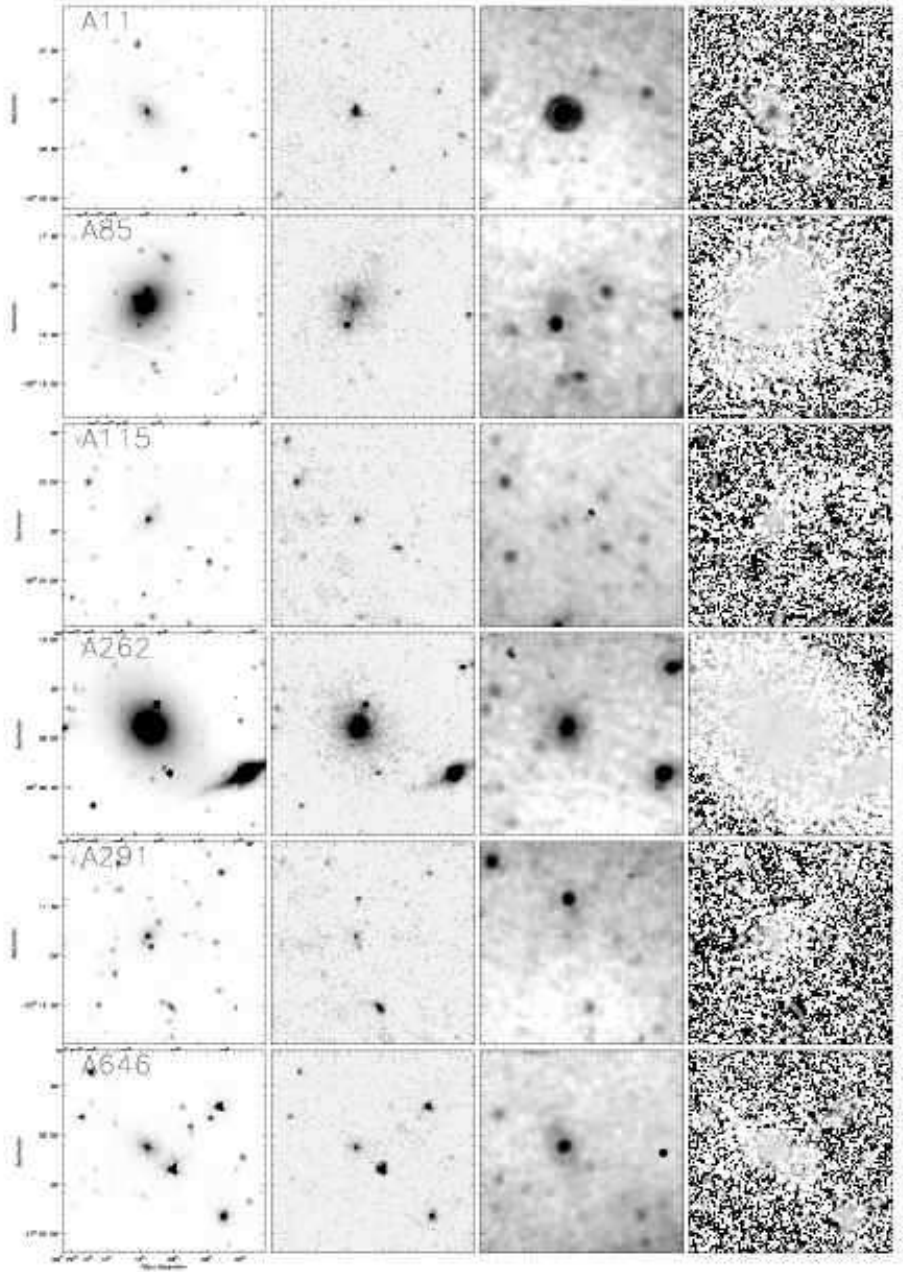}
\caption{Each row shows one brightest cluster galaxy.
From left to right are shown the IRAC
Band 1 (3.6~$\mu$m), the IRAC Band 4 (8~$\mu$m)
and the MIPS Band 1 (24$\mu$m) images.
The rightmost image shows a color map made
from dividing sky subtracted IRAC Band 1 and 4 images in the same region.
The region shown is 2\arcmin$\times$2\arcmin, and approximately
centered on the brightest cluster galaxy.
The color map is on a linear not log grayscale.
Images that contain strong point sources at 24 $\mu$m sometimes exhibit
diffraction rings and spikes (e.g., A1068 and A1664).
Linear features in the images of PKS0745-19, R0338+09 and Z8193 are
due to spikes caused by bright nearby stars.
\label{fig:images}
}
\end{figure*}

\setcounter{figure}{1}

\begin{figure*}
\plotone{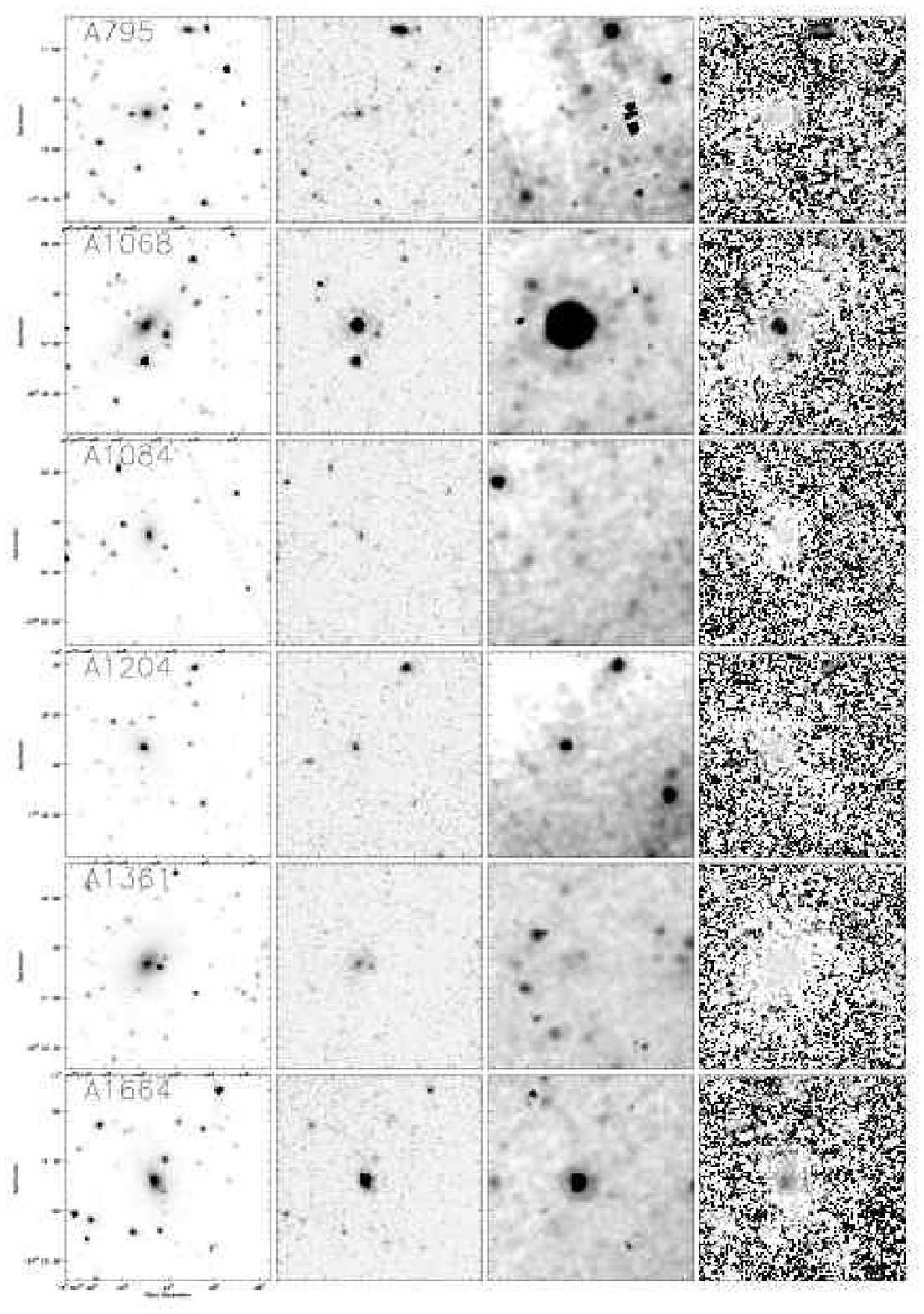}
\caption{continued}
\end{figure*}

\setcounter{figure}{1}
\begin{figure*}
\plotone{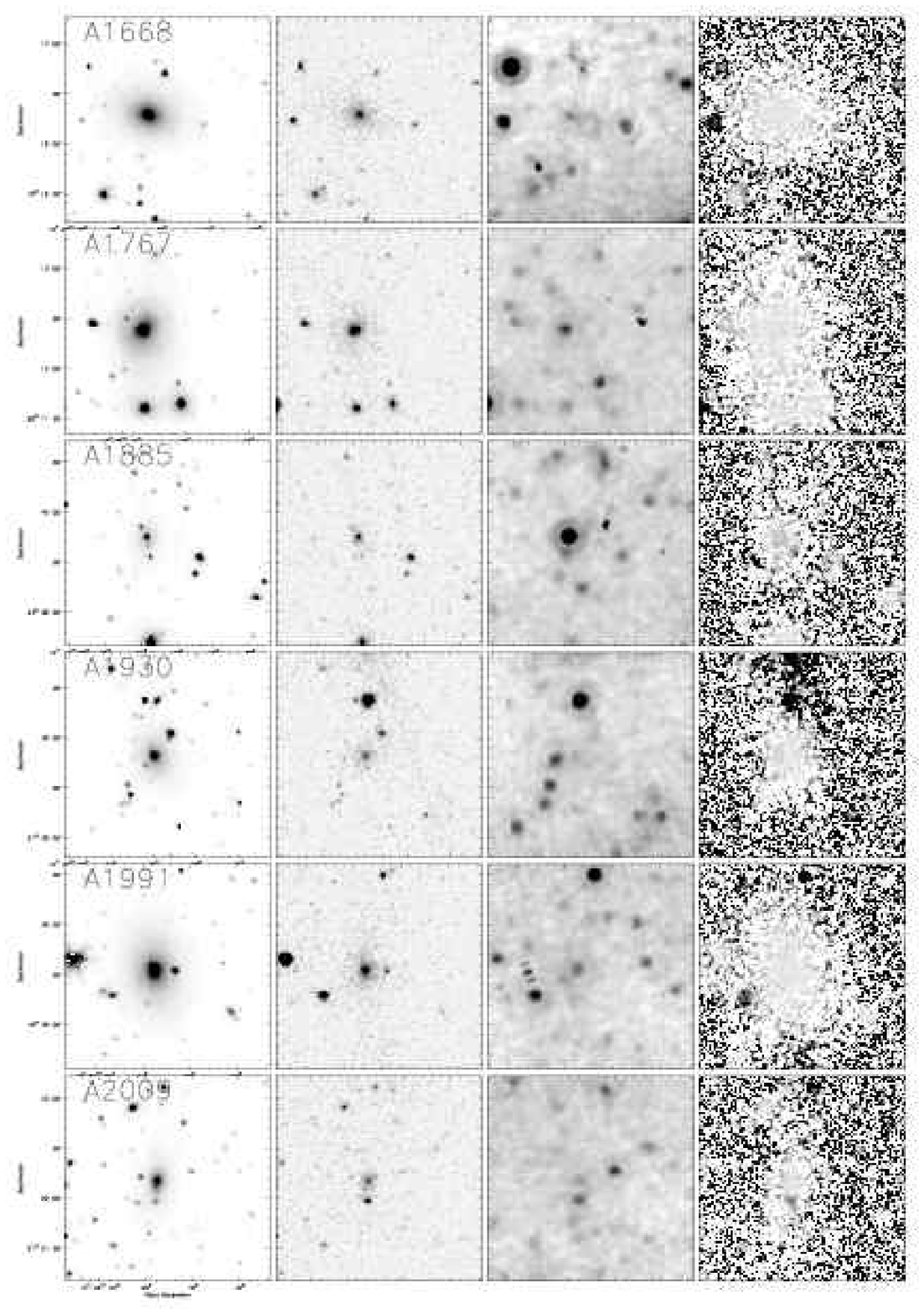}
\caption{continued}
\end{figure*}

\setcounter{figure}{1}
\begin{figure*}
\plotone{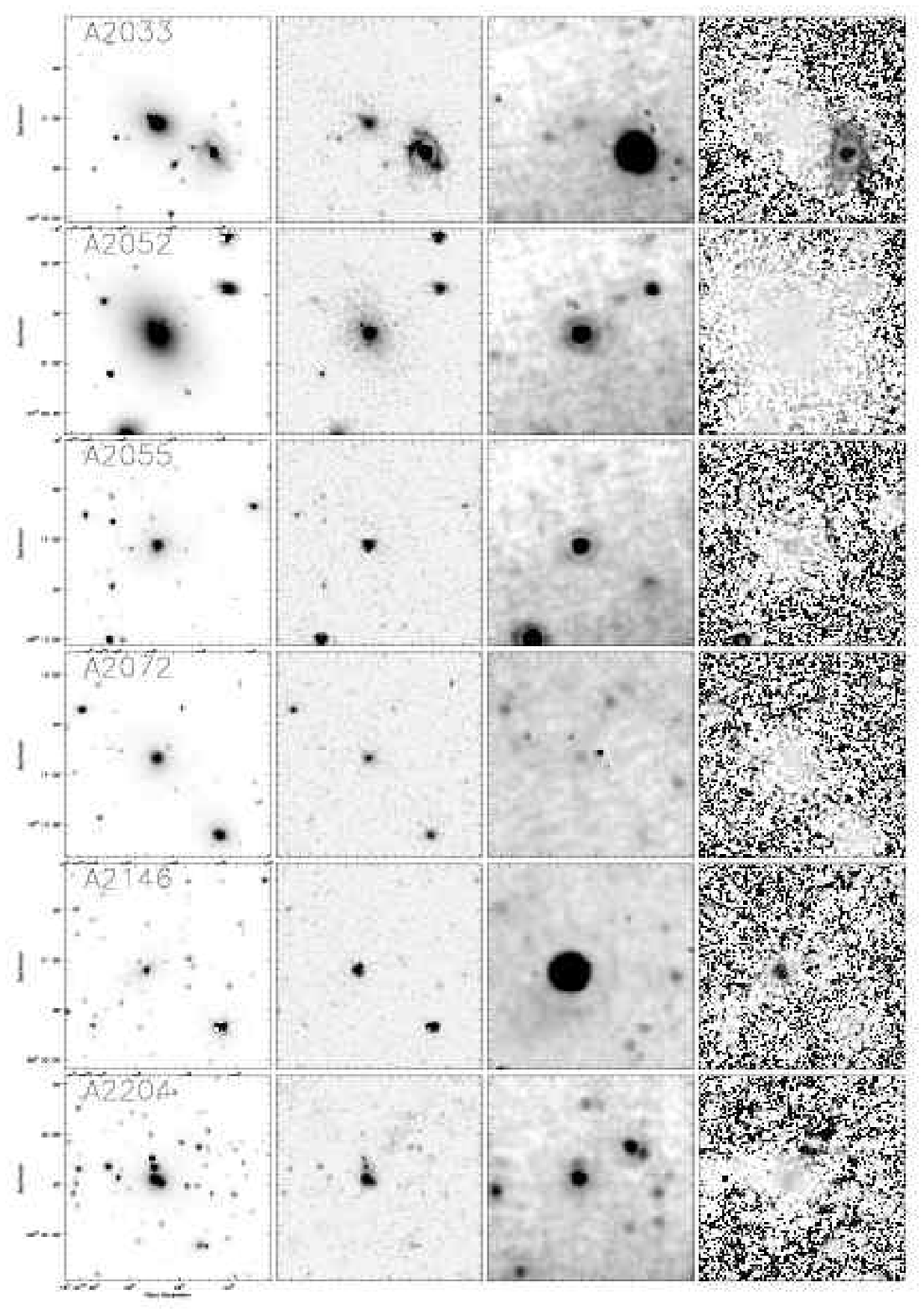}
\caption{continued}
\end{figure*}

\setcounter{figure}{1}
\begin{figure*}
\plotone{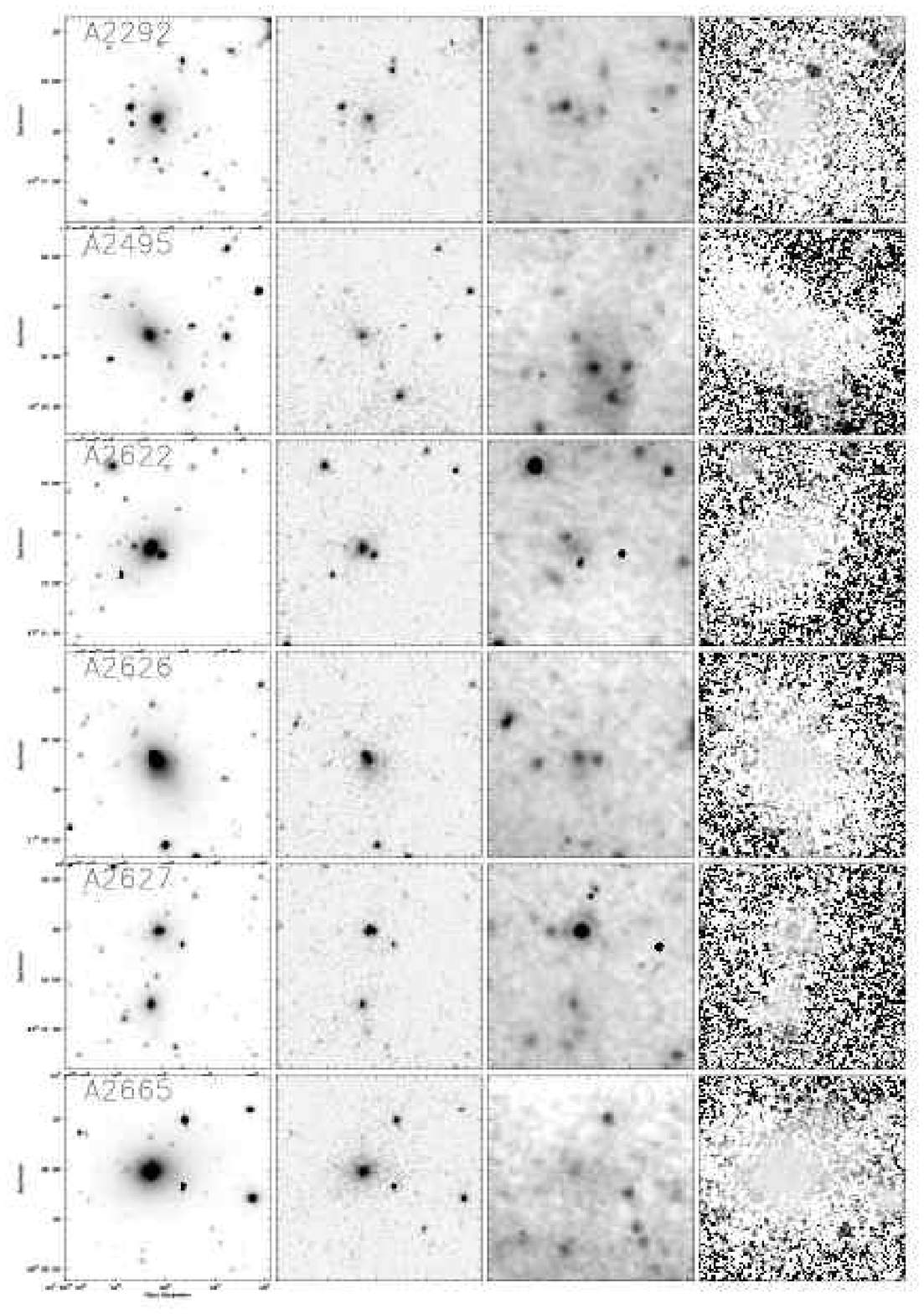}
\caption{continued}
\end{figure*}

\setcounter{figure}{1}
\begin{figure*}
\plotone{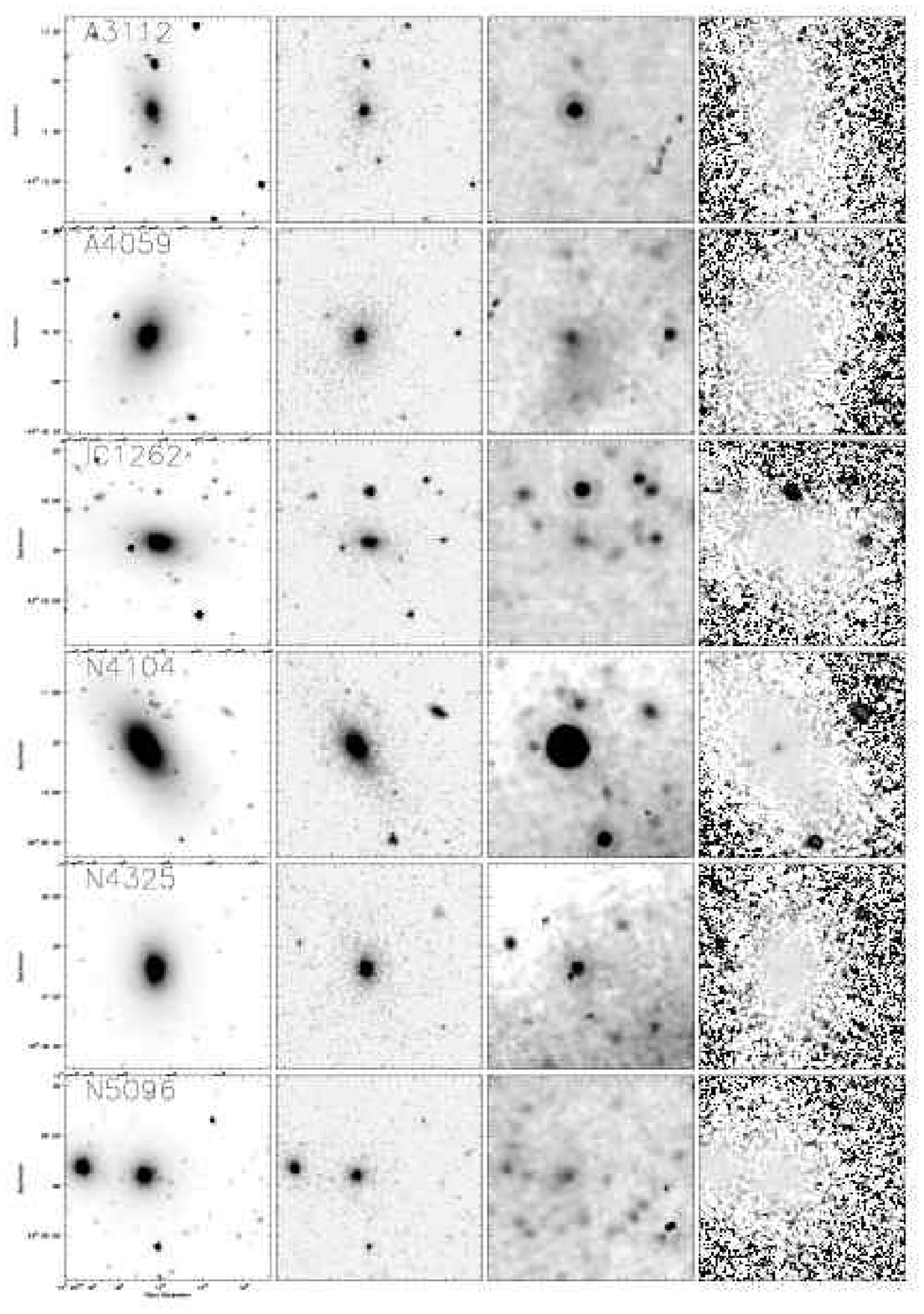}
\caption{continued}
\end{figure*}

\setcounter{figure}{1}
\begin{figure*}
\plotone{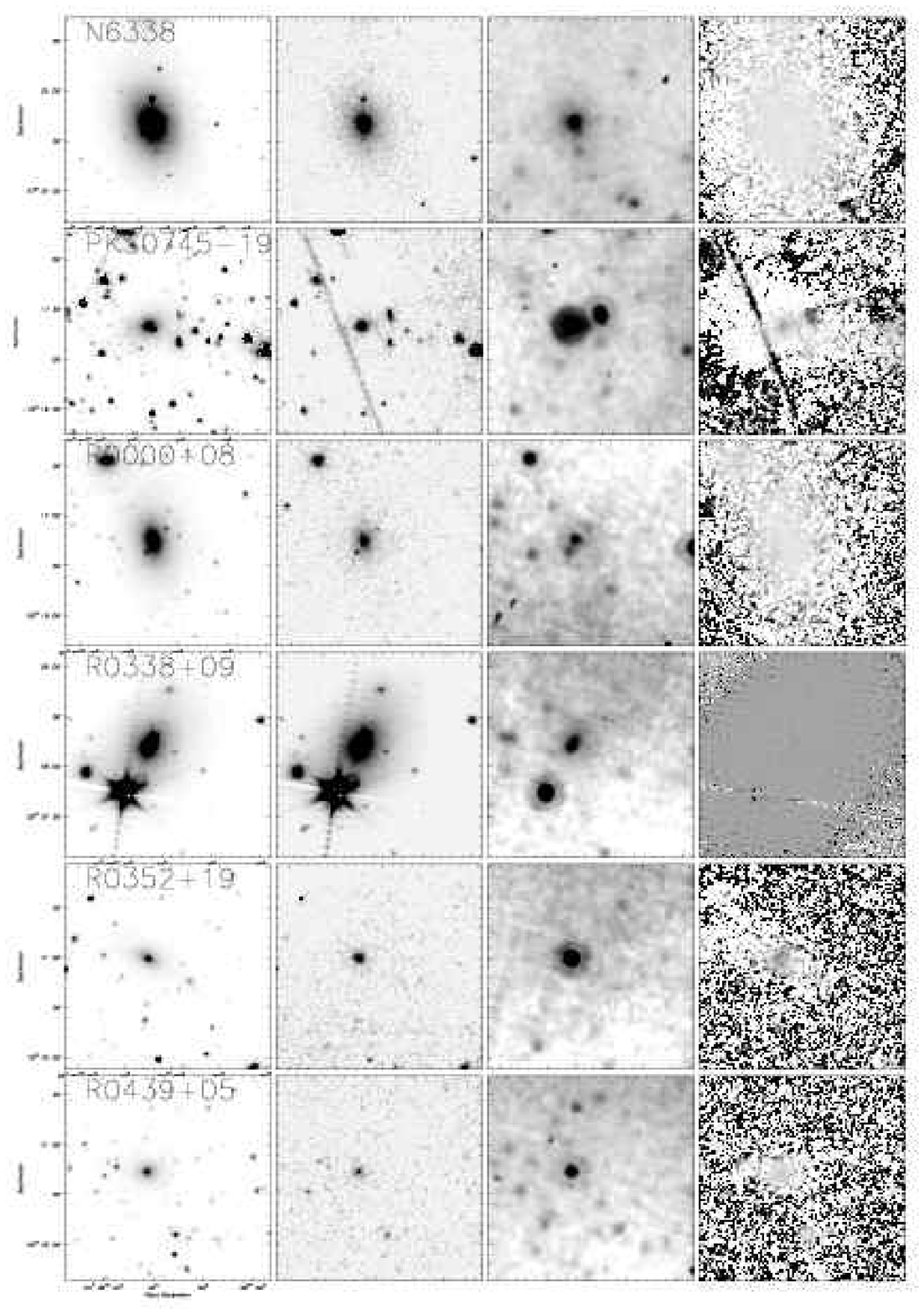}
\caption{continued}
\end{figure*}

\setcounter{figure}{1}
\begin{figure*}
\plotone{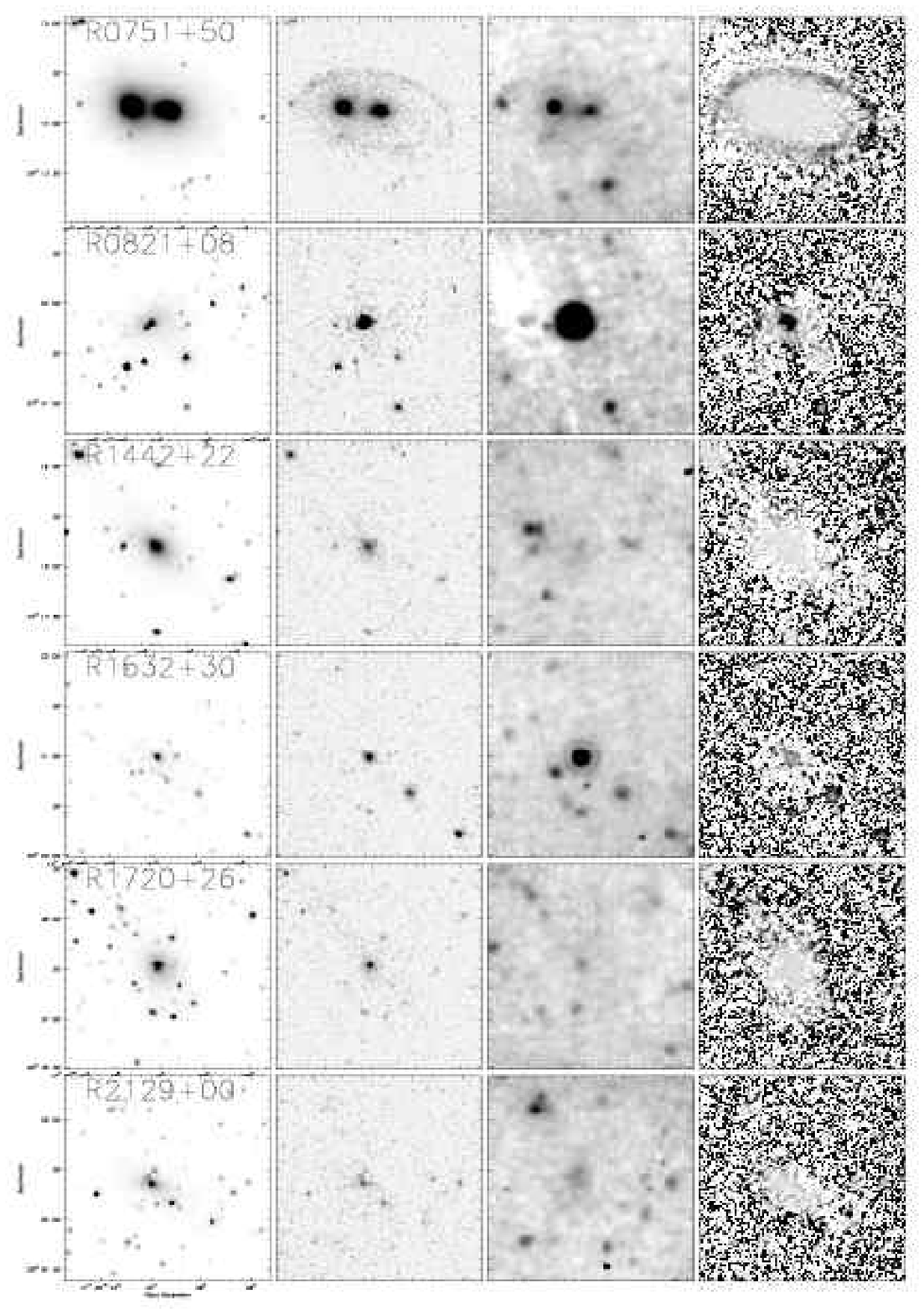}
\caption{continued}
\end{figure*}

\setcounter{figure}{1}
\begin{figure*}
\plotone{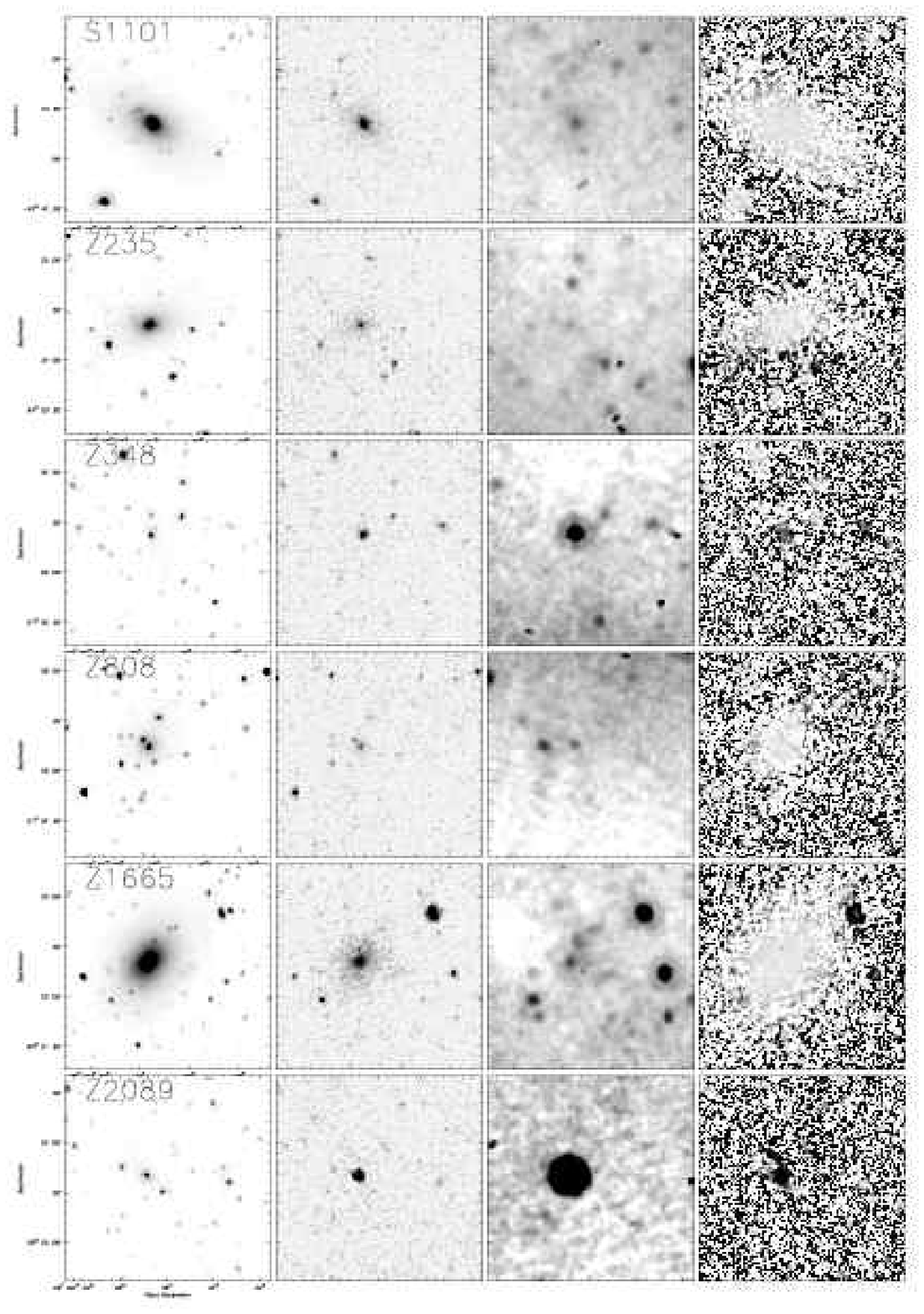}
\caption{continued}
\end{figure*}

\setcounter{figure}{1}
\begin{figure*}
\plotone{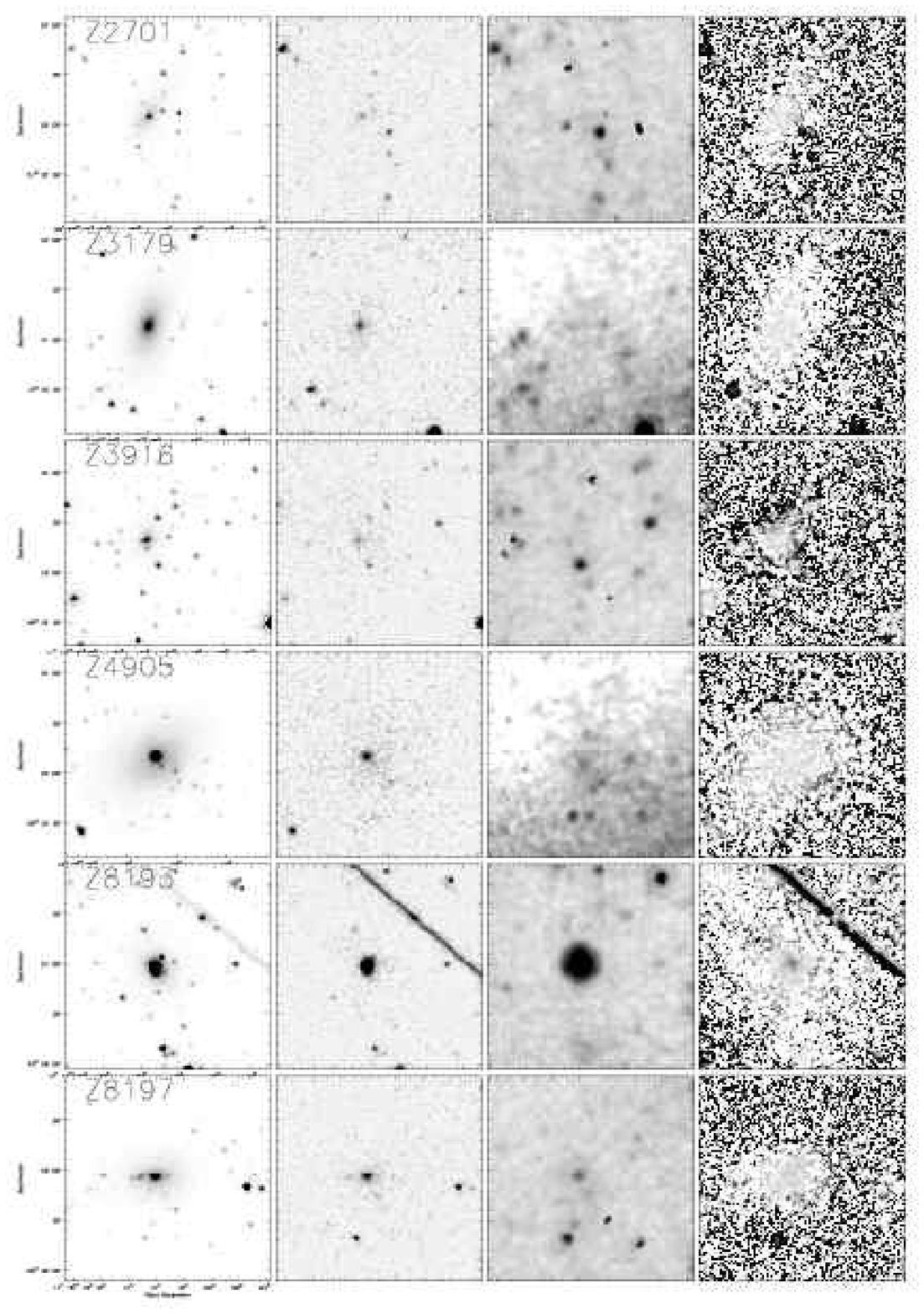}
\caption{continued}
\end{figure*}

\setcounter{figure}{1}
\begin{figure*}
\plotone{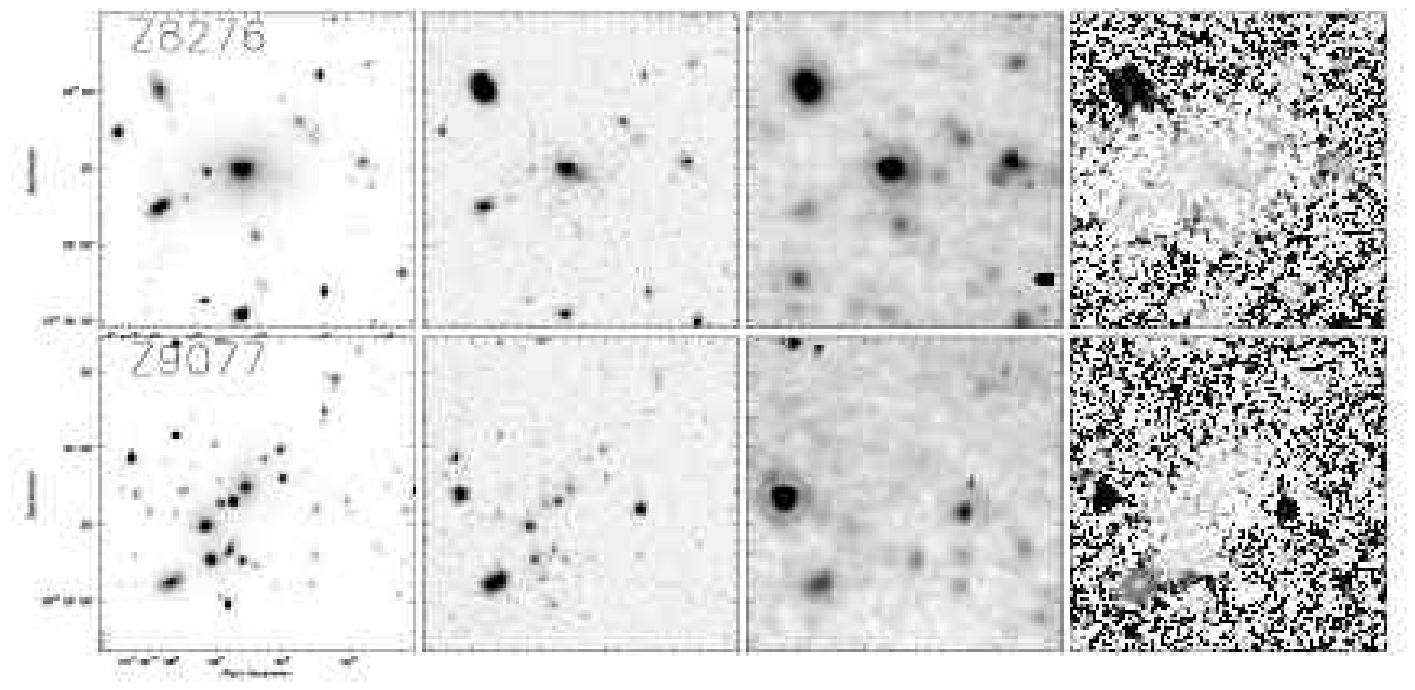}
\caption{continued}
\end{figure*}

\clearpage



\begin{figure*}
\plottwo{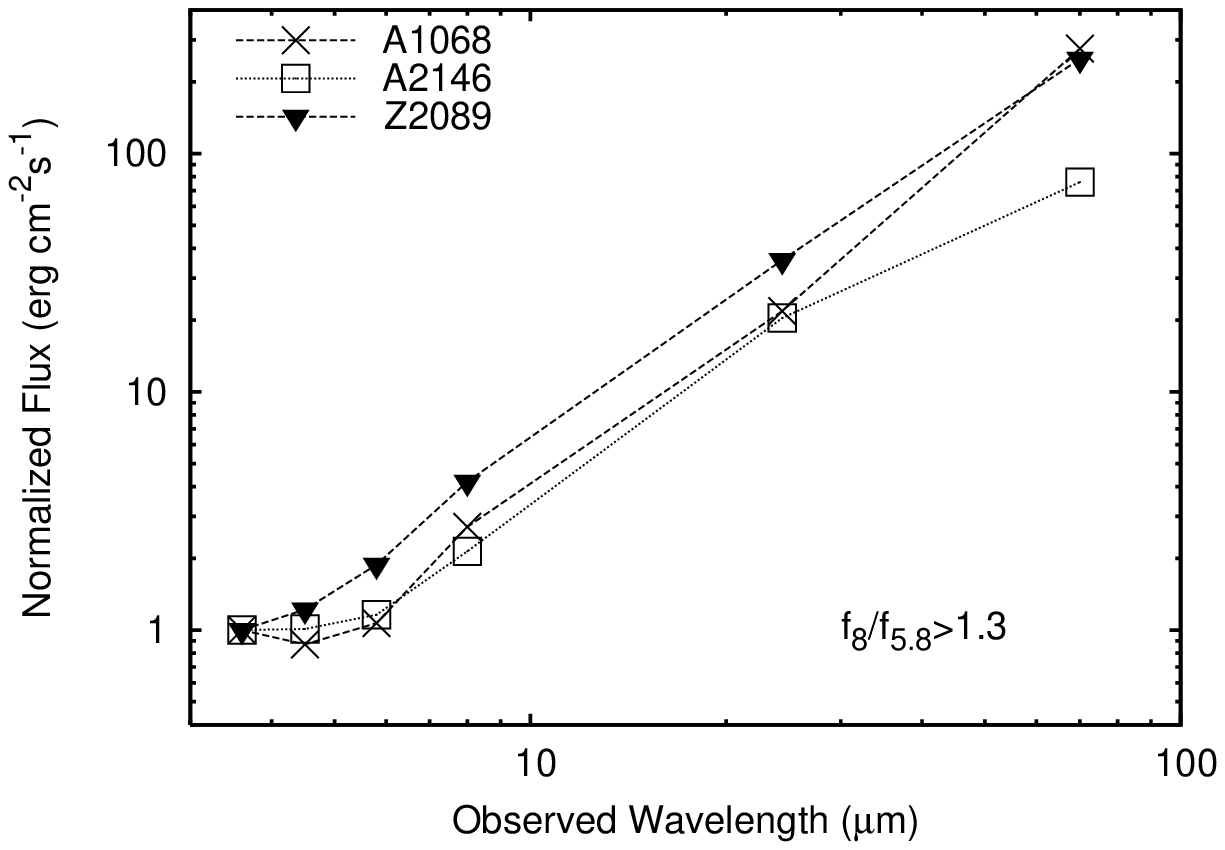}{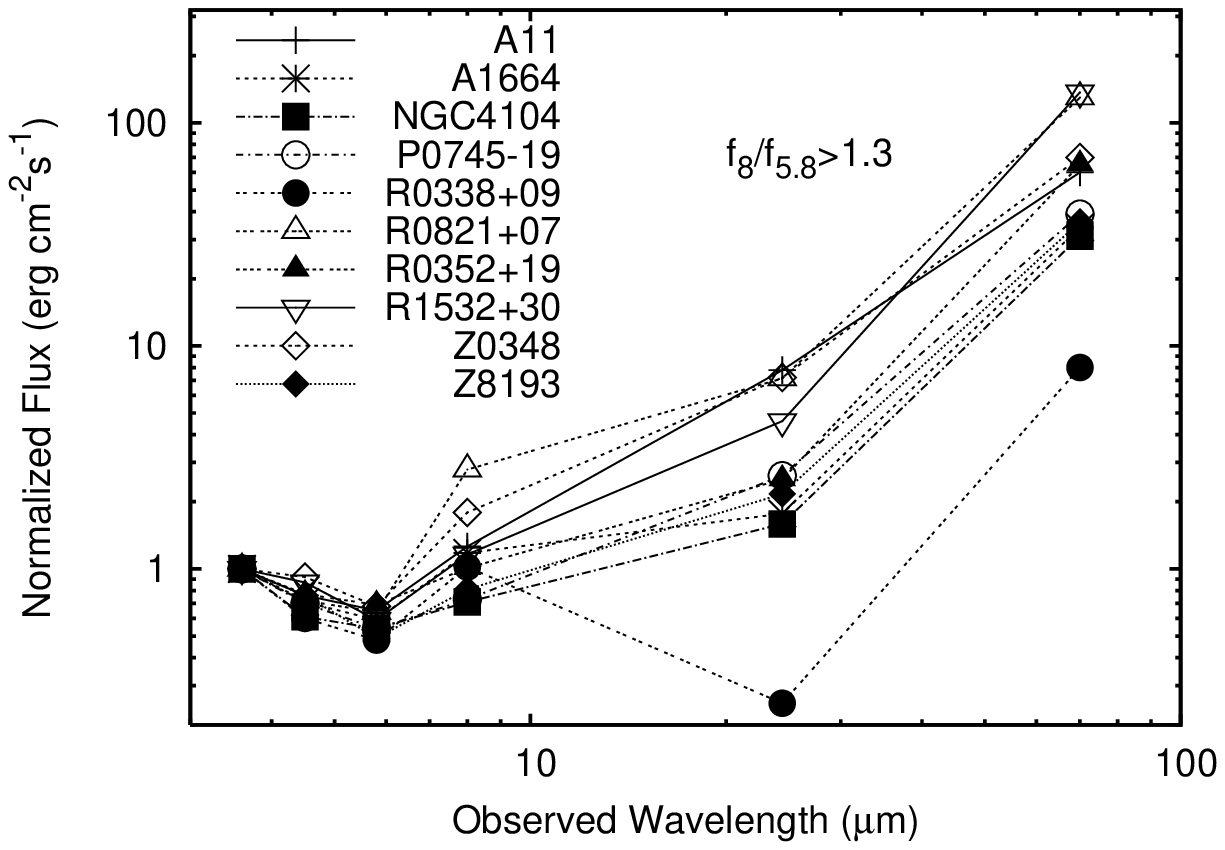}
\plottwo{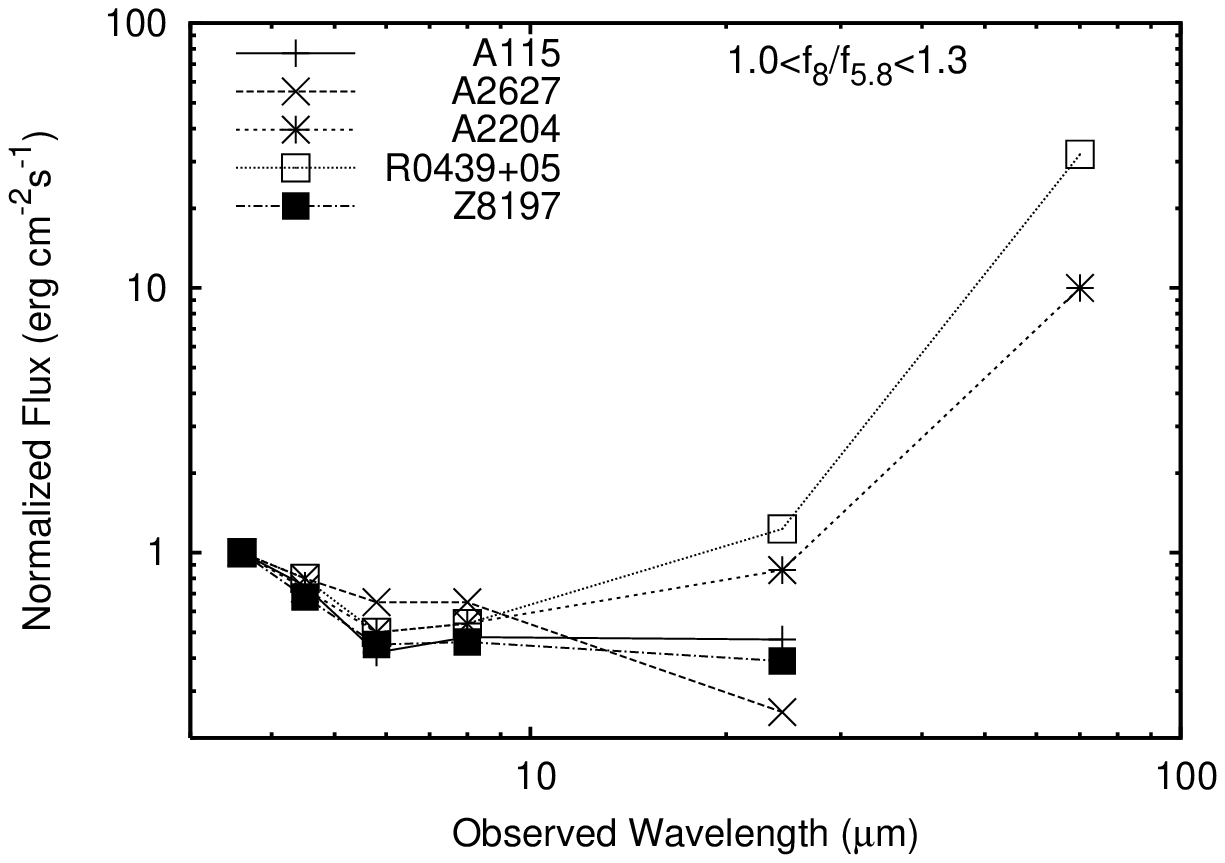}{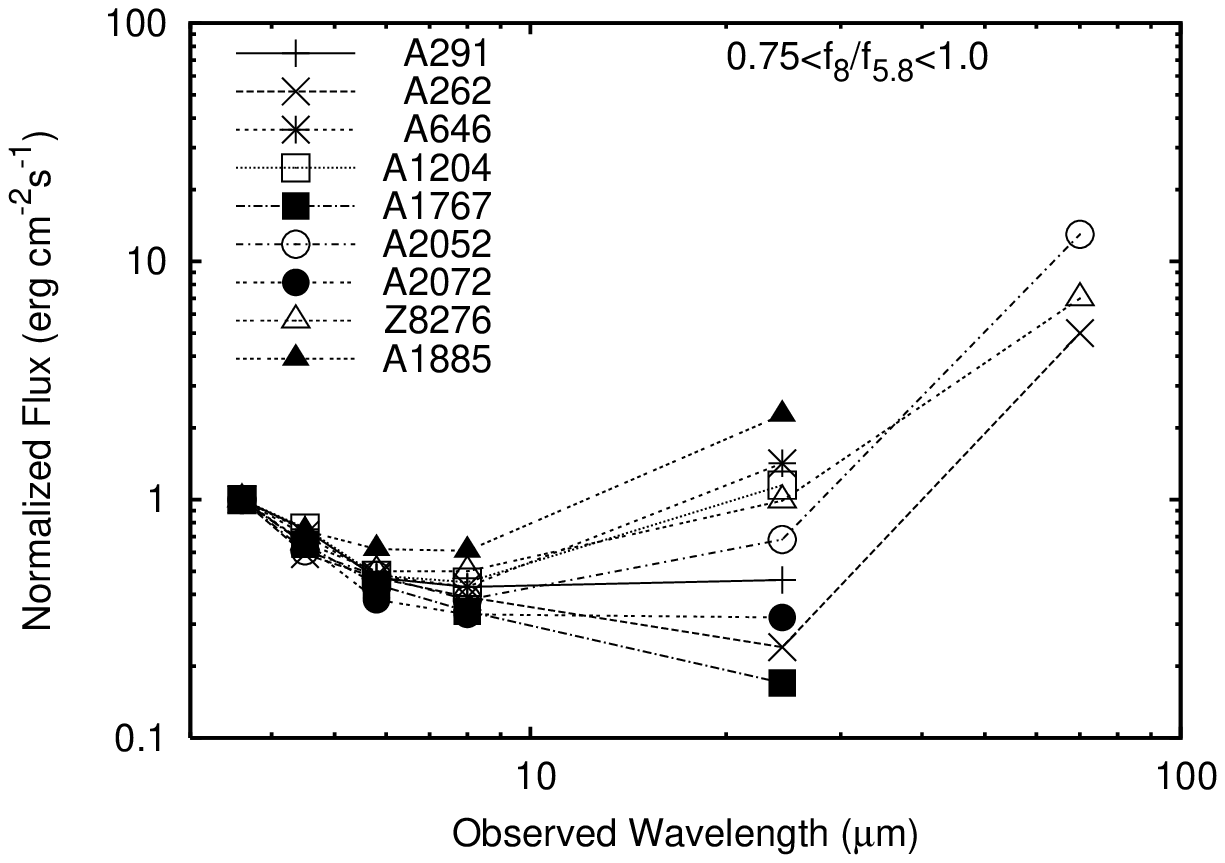}
\caption{Spectral energy distributions of
the brightest cluster galaxies.  We have
grouped them by 5.5/8.0~$\mu$m color.  The fluxes are 
normalized to the 3.6~$\mu$m flux.  
a) These objects are among the reddest objects with
$F_{8 \mu {\rm m}}/F_{5.8 \mu {\rm m}}>1.3$.  They also have unresolved
red nuclei seen in the IRAC color maps.
These three  have high $[$OIII$]$/H$\beta$
ratios suggesting that they host an AGN.  Abell 2146
has a blue 24 to 70 $\mu$m flux ratio suggesting that the dust
is warmer than the others.
b) Additional galaxies with 
$F_{8 \mu {\rm m}}/F_{5.8 \mu {\rm m}}>1.3$.
These galaxies
have strong mid-infrared excesses compared to a quiescent elliptical galaxy 
and all were detected at $70\mu$m.
c) BCGs with $1.0<F_{8 \mu {\rm m}}/F_{5.8 \mu {\rm m}}<1.3$.
d,e) BCGs with $0.75<F_{8 \mu {\rm m}}/F_{5.8 \mu {\rm m}}<1.0$.
f,g) BCGs with $F_{8 \mu {\rm m}}/F_{5.8 \mu {\rm m}}<0.75$.
Most of these are consistent with a quiescent stellar population.  
Objects that are not detected at 70$\mu$ have upper limits on this plot
at 70$\mu$m $\sim 3$--10 times the flux at 3.6$\mu$m.
h) Additional BCGs with $F_{8 \mu {\rm m}}/F_{5.8 \mu {\rm m}}<0.75$
and that are detected at $70\mu$m.
\label{fig:sed}}
\end{figure*}

\setcounter{figure}{2}
\begin{figure*}
\plottwo{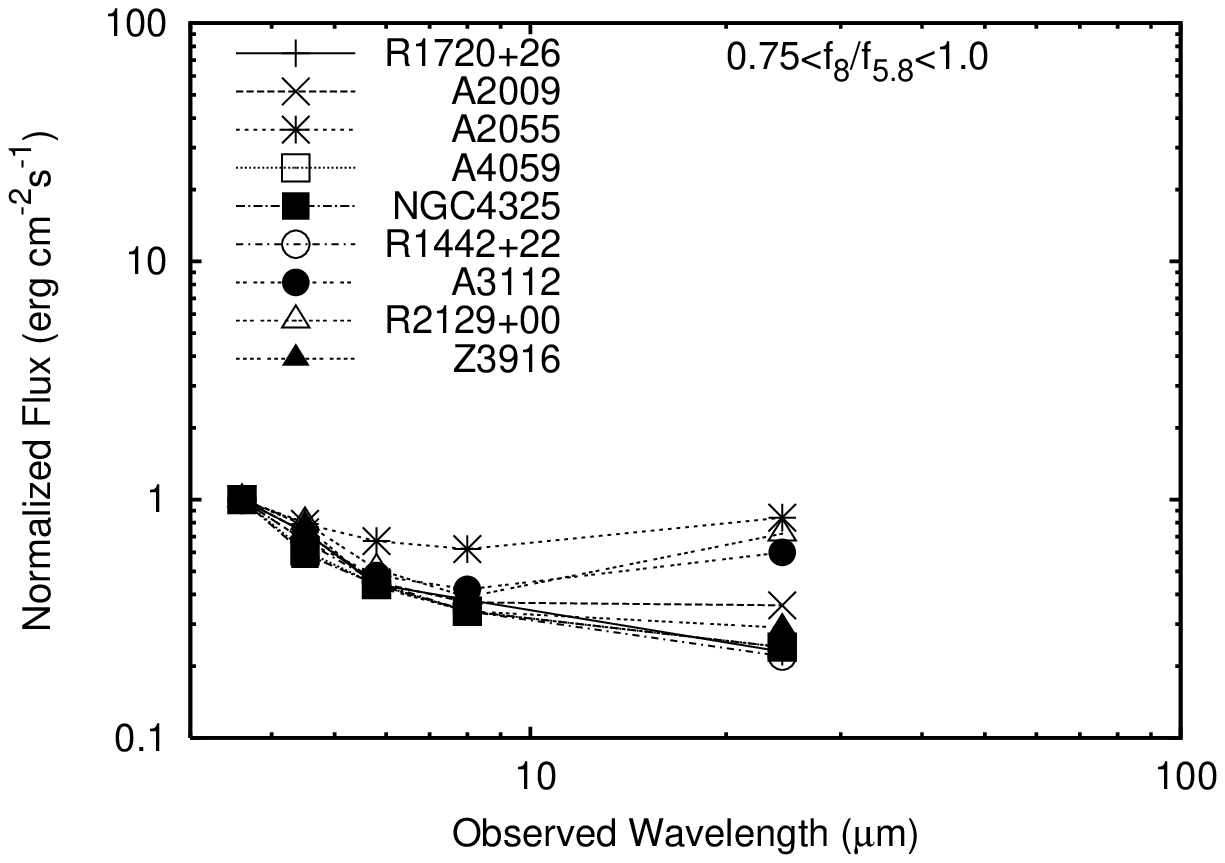}{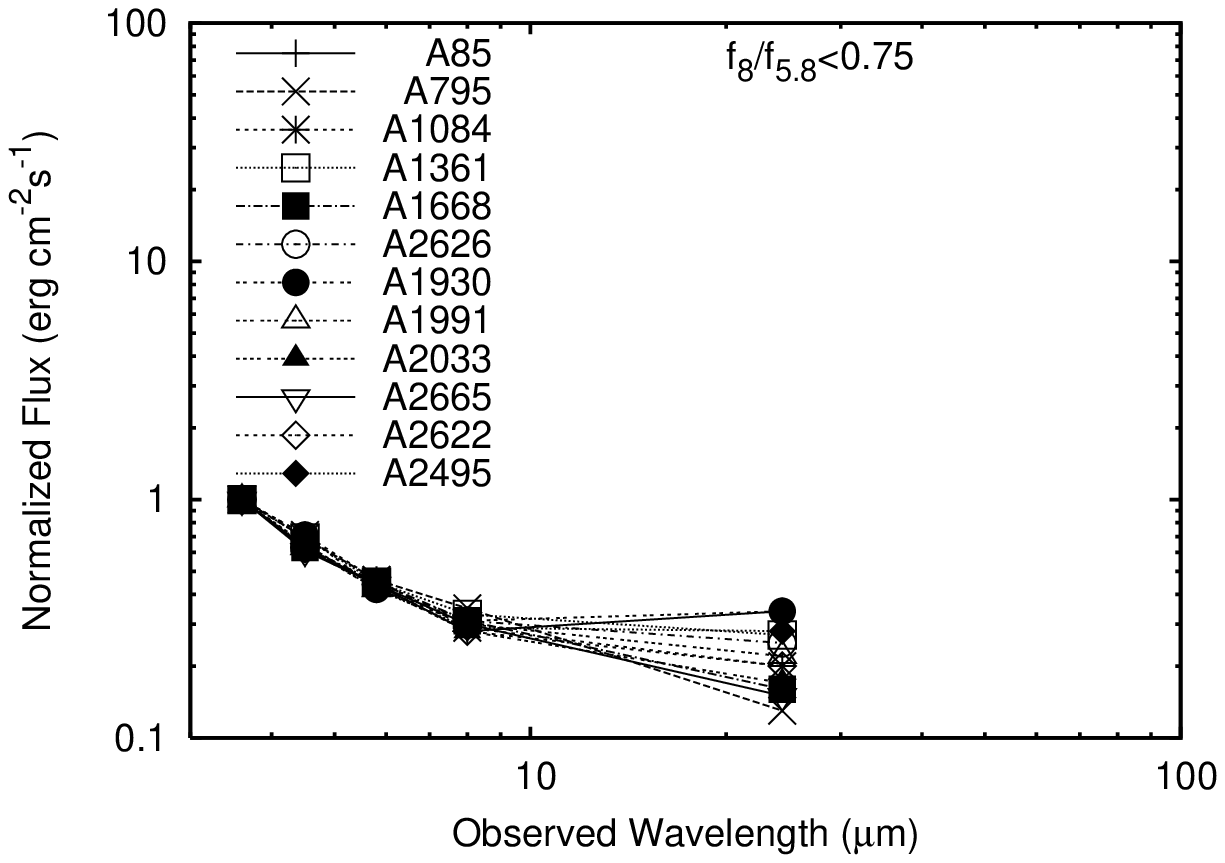}
\plottwo{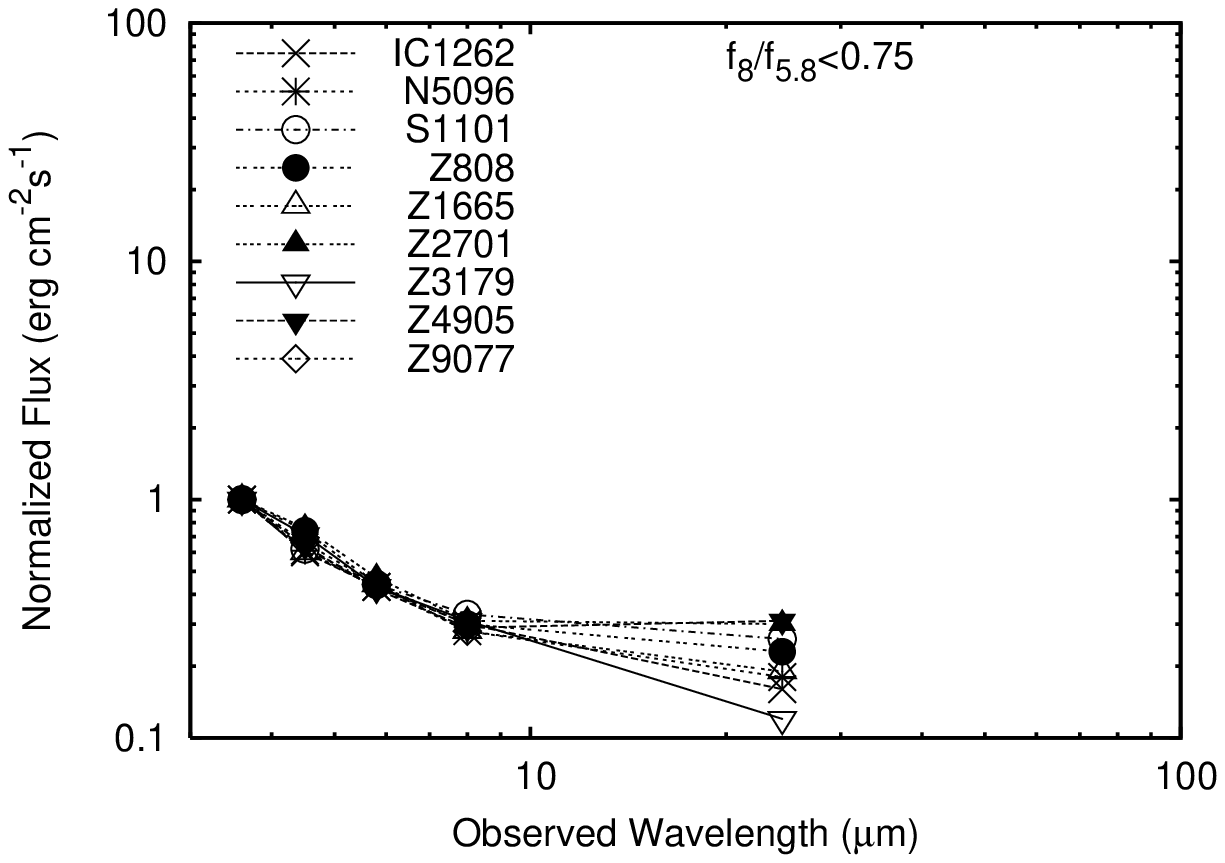}{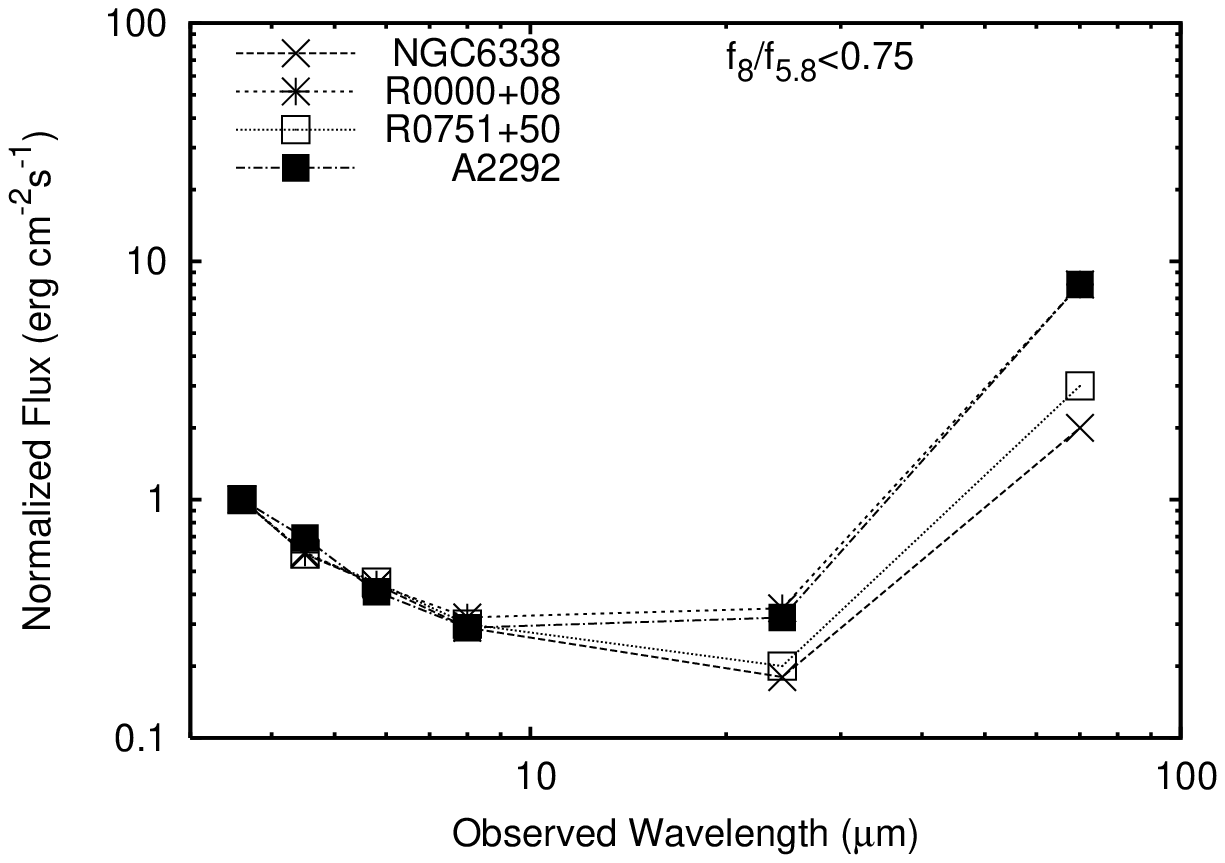}
\caption{continued}
\end{figure*}

\begin{figure*}
\plotone{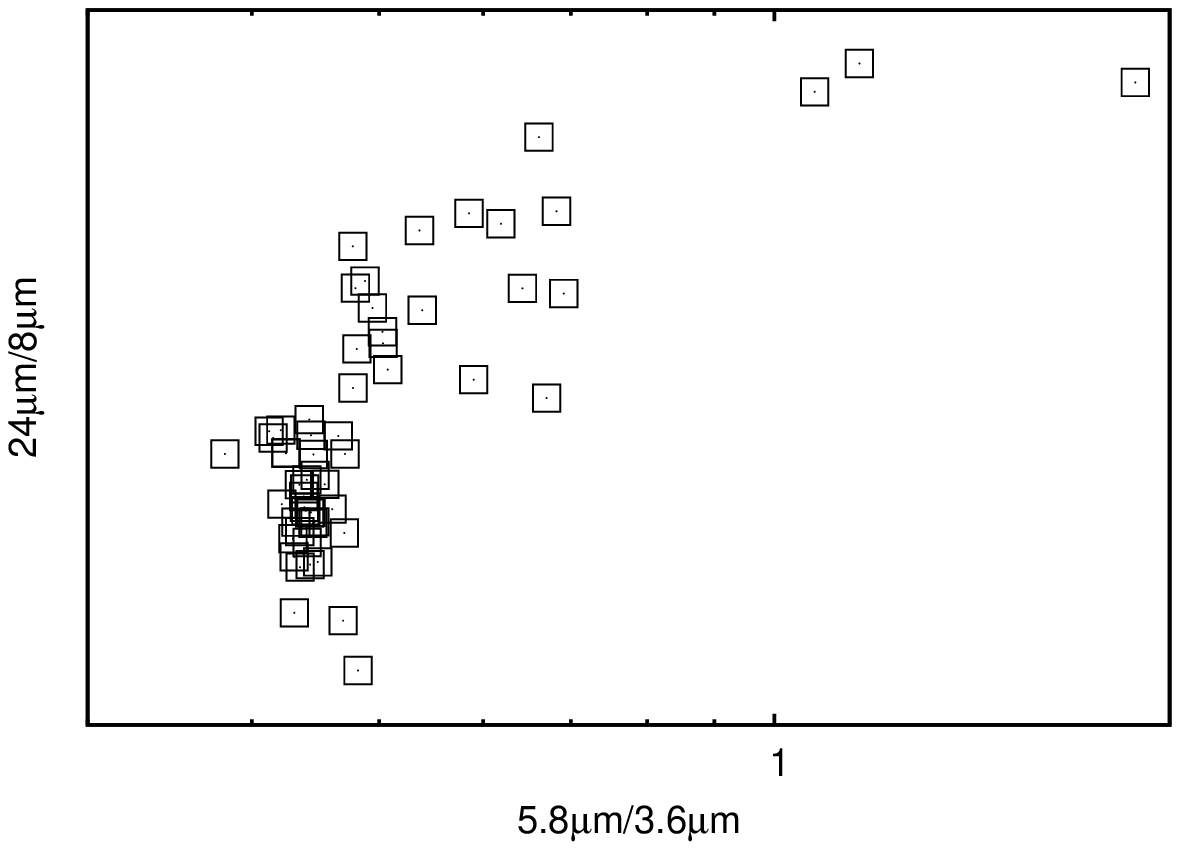}
\caption{Infrared color-color plot.  Each axis is the ratio of the fluxes
at the listed wavelengths. 
Clusters with a 5.8/3.6~$\mu$m flux ratio less than 0.5 and a 24/8~$\mu$m
flux ratio less than 2 are quiescent.
The two color ratios are correlated but with larger scatter in
the 24/8~$\mu$m ratio than at shorter wavelengths.
Objects with flux ratios $F_{5.8}/F_{3.6} \ga 0.5$ or $F_{24}/F_8 \ga 1$
have infrared excesses.
The three objects at the top right of the plot are 
Abell 1068, Abell 2146 and Z2089
with unresolved red nuclear sources in the IRAC bands
indicating the presence of hot dust.
These three also have high measured $[$OIII$]$/H$\beta$ ratios
and so probably host a dusty AGN.
\label{fig:colorcolor}}
\end{figure*}


\begin{thebibliography}{}



\bibitem[Allen(1995)]{allen95}
Allen, S. W. 1995, MNRAS, 276, 947

\bibitem[Allen(2000)]{allen00}
Allen, S. W. 2000, MNRAS, 315, 269

\bibitem[Alonso-Herrero et al.(2003)]{alonso03}
Alonso-Herrero, A., Quillen, A. C., Rieke, G. H., Ivanov, V.D., \&
Efstathiou, A.  2003, AJ, 126, 81	

\bibitem[Armus et al.(2007)]{armus07}
Armus, L., et al. 2007, ApJ, 656, 148

\bibitem[Becker et al.(1995)]{becker95}
Becker, R. H., White, R. L., \& Helfand, D. J. 1995,
ApJ, 450, 559





\bibitem[Birzan et al.(2004)]{birzan04}
Birzan, L., Rafferty, D. A., McNamara, B. R., Wise, M. W., \& Nulsen, P. E. J.	
 2004, ApJ, 607, 800


\bibitem[B\"ohringer et al.(2004)]{bohringer04}
B\"ohringer, H., et al. 2004, A\&A, 425, 367

\bibitem[Bregman et al.(2006)]{bregman06}
Bregman, J. N., Fabian, A. C., Miller, E. D., \& Irwin, J. A.	
 2006, ApJ, 642, 746


\bibitem[Brinkmann et al.(1995)]{brinkmann95}
Brinkmann, W., Siebert, J., Reich, W., Fuerst, E., Reich, P., Voges, W., Truemper, J., 
\& Wielebinski, R.  1995, A\&AS, 109, 147


\bibitem[Cardiel et al.(1998)]{cardiel98}
Cardiel, N., Gorgas, J., \& Aragon-Salamanca, A.
1998, Ap\&SS, 263, 83	

\bibitem[Condon et al.(1998)]{condon98}
Condon, J. J., et al. 1998, AJ, 115, 1693


\bibitem[Crawford et al.(1999)]{crawford99}
Crawford, C. S., Allen, S. W., Ebeling, H., Edge, A. C., \& Fabian, A. C.
 1999, MNRAS, 306, 857

\bibitem[David et al.(1999)]{david99}
David, L.P., Forman, W., \& Jones, C. 1999, ApJ, 519, 533


\bibitem[Donahue et al.(2007)]{donahue07}
Donahue, M., Baum, S., Cote, P., Ferrarese, L., Goudfrooij, P.,
Jordan, A., Macchetto, D., Malhotra, S., O'Dea, C. P.,
Pringle, J., Rhoads, J., Sparks, W., \& Voit, G. M.
2007, ApJ, in press

\bibitem[Dunn \& Fabian(2006)]{dunn06}
Dunn, R. J. H. \& Fabian A. C.  2006, MNRAS, 373, 959

\bibitem[Ebeling et al.(1998)]{ebeling98}
Ebeling, H., Edge, A. C., B\"ohringer, H., Allen, S. W.,
Crawford, C. S., Fabian, A. C., Voges, W., \& Huchra, J. P. 1998,
MNRAS, 301, 881


\bibitem[Ebeling et al.(2000)]{ebeling00}
Ebeling, H., Edge, A. C., Allen, S. W., Crawford, C. S., Fabian, A. C., \&
Huchra, J. P.   2000, MNRAS, 318, 333

\bibitem[Edge et al.(1990)]{edge90}
Edge, A. C., Stewart, G. C., Fabian, A. C., \& Arnaud, K. A.
 1990, MNRAS, 245, 559

\bibitem[Edge et al.(1992)]{edge92}
Edge, A. C., Stewart, G. C., \& Fabian, A. C. 1992, MNRAS, 258, 177

\bibitem[Edge et al.(2002)]{edge02}
Edge, A. C., Wilman, R. J., Johnstone, R. M., Crawford, C. S.,
Fabian, A. C., \& Allen, S. W.	2002, MNRAS, 337, 49	
 

\bibitem[Edge(2001)]{edge01}
Edge, A. C. 2001, MNRAS, 328, 762


\bibitem[Egami et al.(2006)]{egami06}
Egami, E., et al. 2006, ApJ, 647, 922



\bibitem[Elbaz et al.(2002)]{elbaz02}
Elbaz, D., Cesarsky, C. J., Chanial, P., Aussel, H., Franceschini, A.,
Fadda, D., \& Chary, R. R. 2002, A\&A, 384, 848



\bibitem[Fabian \& Crawford(1995)]{fabian95}
Fabian, A. C., \& Crawford, C. S. 1995, MNRAS, 274, L63

\bibitem[Fazio et al.(2004)]{fazio04}
Fazio, G. G., et al. 2004, ApJS, 154, 10

\bibitem[Hatch et al.(2007)]{hatch07}
Hatch, N. A., Crawford, C. S., \& Fabian, A. C.	
2007, MNRAS, 380, 33	


\bibitem[Heckman (1981)]{heckman81}
Heckman, T. M. 1981, ApJ, 250, L59

\bibitem[Hicks \& Mushotzky(2005)]{hicks05}
Hicks, A. K., \& Mushotzky, R.	2005, ApJ, 635, L9	

\bibitem[Hu, Cowie \& Wang (1985)]{Hu85}
Hu, E. M., Cowie, L. L., \& Wang, Z. 1985, ApJS, 59, 447

\bibitem[Jaffe et al.(2005)]{jaffe05}
Jaffe, W., Bremer, M. N., \& Baker, K.  2005, MNRAS, 360, 748


\bibitem[Jones et al.(2004)]{jones04}
Jones, D. H., et al. 2004, MNRAS 355, 747

\bibitem[Jones et al.(2006)]{jones06}
Jones, D. H., Saunders, W., Read, M. A., \& Colless, M. 2005, PASA, 22, 277 

\bibitem[Katgert et al.(1998)]{katgert98}
Katgert, P., Mazure, A., den Hartog, R., Adami, C., Biviano, A., \& Perea, J.	
1998, A\&AS, 129, 399



\bibitem[Makovoz \& Khan(2005)]{makovoz05}
Makovoz, D., \& Khan, I. 2005, in ASP Conf. Ser. 347, 
Astronomical Data Analysis 
Software and Systems XIV, ed. P. Shopbell, M. Britton, \& R. Ebert 
(San Francisco: ASP), 81


\bibitem[McNamara et al.(2004)]{mcnamara04b}
McNamara, B. R., Wise, M. W., \& Murray, S. S. 2004, ApJ, 601, 173

\bibitem[McNamara \& O'Connell(1989)]{mcnamara89}
McNamara, B. R., \& O'Connell, R. W. 1989, AJ, 98, 2018

\bibitem[McNamara(2004)]{mcnamara04}
Star Formation in Cluster Cooling Flows,
McNamara, B. R., Proceedings of The Riddle of Cooling Flows in Galaxies
and Clusters of Galaxies, held in Charlottesville, VA, May 31 -
June 4, 2003, Eds. T. Reiprich, J. Kempner, and N. Soker., page 177

\bibitem[McNamara \& Nulsen(2007)]{mcnamara07}
McNamara, B. R. \& Nulsen, P. E. J. 2007, ARA\&A, 45, 117 





\bibitem[O'Dea et al.(2007)]{odea07}
O'Dea, C. P., et al. 2007, in preparation

\bibitem[Owen et al.(1995)]{owen95}
Owen, F. N., Ledlow, M. J., \& Keel, W. C. 1995, AJ, 109, 14O	

\bibitem[Peres et al.(1998)]{peres98}
Peres, C. B., Fabian, A. C., Edge, A. C., Allen, S. W., Johnstone, R. M.,
\& White, D. A.	1998, MNRAS, 298, 416


\bibitem[Peterson \& Fabian(2006)]{peterson06}
Peterson, J. R., \& Fabian, A. C. 2006, Phys Rep, 427, 1

\bibitem[Rafferty et al.(2006)]{rafferty06}
Rafferty, D. A., McNamara, B. R., Nulsen, P. E. J., \& Wise, M. W.
2006, ApJ, 652, 216

\bibitem[Rieke et al.(2004)]{rieke04}
Rieke, G. H., et al. 2004, ApJS, 154, 25

\bibitem[Rush, Malkan \& Spinoglio (1993)]{rush93}
Rush, B., Malkan, M. A., \& Spinoglio, L. 1993, ApJS, 89, 1



\bibitem[Sanders et al.(1988)]{sanders88}
Sanders, D. B., Soifer, B. T., Elias J. H., Neugebauer, G.,
\& Matthews, K. 1988, ApJ, 328, L35

\bibitem[Salom\'e \& Combes(2003)]{salome03}
Salom\'e, P., \& Combes, F. 2003, A\&A, 412, 657

\bibitem[Schweitzer et al.(2006)]{schweitzer06}
Schweitzer, M., et al., 2006, ApJ, 649, 79

\bibitem[Spinoglio et al.(1995)]{spinoglio}
Spinoglio, L., Malkan, M.~A., Rush, B., Carrasco, L., \& Recillas-Cruz, E.\
1995, ApJ, 453, 616

\bibitem[Voit \& Donahue(1997)]{voit97}
Voit, G. M., \& Donahue, M. 1997, ApJ, 482, 242

\bibitem[White et al.(1997)]{white97}
White, R. L., Becker, R. H., Helfand, D. J., \&
Gregg, M. D. 1997, ApJ, 475, 479

\bibitem[Wilman et al.(2006)]{wilman06}
Wilman, R. J., Edge, A. C., \& Swinbank, A. M. 2006, MNRAS, 371, 93



\end{thebibliography}
\end{document}